\documentclass[12pt]{article}
\usepackage{graphics,graphicx}
\usepackage{amssymb,epsfig,amsmath,euscript,array}
\usepackage{cite}
\usepackage{axodraw}
\usepackage{pstricks}
\usepackage{color}

\usepackage{subfigure}
\usepackage{tikz}
\usetikzlibrary{shapes,arrows,positioning}

\makeatletter
\@addtoreset{equation}{section}
\makeatother

\newcounter{multieqs}

\newcommand{\spaa[1]}{\langle #1 \rangle}

\newcommand{\be}{\begin{equation}}
\newcommand{\ee}{\end{equation}}

\newcommand{\bm}[1]{\mbox{\boldmath $#1$}}

\newcommand{\kslash}{k \!\!\! / }

\newcommand{\lslash}{l \!\! / }
\newcommand{\Pslash}{P \!\!\!\! / }

\newcommand{\islash}{i \!\!\! / }
\newcommand{\jslash}{j \!\!\! / }
\newcommand{\aslash}{a \!\!\! / }
\newcommand{\bslash}{{b \hspace{-6pt} \slash} }

\newcommand{\onslash}{1 \!\!\! / }
\newcommand{\twslash}{2 \!\!\!/ }
\newcommand{\thslash}{3 \!\!\!/ }
\newcommand{\foslash}{4 \!\!\! / }
\newcommand{\fislash}{5 \!\!\! / }

\newcommand{\mslash}{m \!\!\! / }

\def\bd{\begin{document}}
\def\ed{\end{document}}
\def\nn{\nonumber}
\def\bea{\begin{eqnarray}}
\def\eea{\end{eqnarray}}

\def\ab{(ijab)}
\def\ba{(ijba)}
\def\ijab{{\tr}_{+}(\islash\, \jslash\, \aslash \, \bslash)}
\def\ijba{{\tr}_{+}(\islash\, \jslash\, \bslash \, \aslash)}
\def\ijaP{{\tr}_{+}(\islash\, \jslash\, \aslash \, \Pslash)}
\def\ijPLa{{\tr}_{+}(\islash\, \jslash\, \Pslash_L \, \aslash)}
\def\ijaPL{{\tr}_{+}(\islash\, \jslash\, \aslash \, \Pslash_L)}
\def\ijPLza{{\tr}_{+}(\islash\, \jslash\, \Pslash_{L;z} \, \aslash)}
\def\ijaPLz{{\tr}_{+}(\islash\, \jslash\, \aslash \, \Pslash_{L;z})}
\def\ijPa{{\tr}_{+}(\islash\, \jslash\, \Pslash \, \aslash)}
\def\iaPb{{\tr}_{+}(\islash\, \aslash\, \Pslash \, \bslash)}
\def\ibPa{{\tr}_{+}(\islash\, \bslash\, \Pslash \, \aslash)}
\def\ijPmu{{\tr}_{+}(\islash\, \jslash\, \Pslash \, \mu)}
\def\ibmuP{{\tr}_{+}(\islash\, \bslash\, \mu \, \Pslash)}
\def\ibmua{{\tr}_{+}(\islash\, \bslash\, \mu \, \aslash)}
\def\iamub{{\tr}_{+}(\islash\, \aslash\, \mu \, \bslash)}
\def\jaPb{{\tr}_{+}(\jslash\, \aslash\, \Pslash \, \bslash)}
\def\ijmuP{{\tr}_{+}(\islash\, \jslash\, \mu \, \Pslash)}
\def\ijmum{{\tr}_{+}(\islash\, \jslash\, \mu \, \mslash)}
\def\ijmmu{{\tr}_{+}(\islash\, \jslash\, \mslash \, \mu)}
\def\ijmP{{\tr}_{+}(\islash\, \jslash\, \mslash \, \Pslash)}
\def\iabP{{\tr}_{+}(\islash\, \aslash\, \bslash \, \Pslash)}
\def\ijbP{{\tr}_{+}(\islash\, \jslash\, \bslash \, \Pslash)}
\def\jbPa{{\tr}_{+}(\jslash\, \bslash\, \Pslash \, \aslash)}
\def\ijPb{{\tr}_{+}(\islash\, \jslash\, \Pslash \, \bslash)}
\def\jbmua{{\tr}_{+}(\jslash\, \bslash\, \mu \, \aslash)}

\def\loablt{ {\tr}_{+}(\lslash_1\, \aslash \, \bslash\, \lslash_2)}

\def\ijlolt{{\tr}_{+}(\islash\, \jslash\, \lslash_1 \, \lslash_2)}
\def\ijltlo{{\tr}_{+}(\islash\, \jslash\, \lslash_2 \, \lslash_1)}
\def\ibloa{{\tr}_{+}(\islash\, \bslash\, \lslash_1 \, \aslash)}
\def\jaltb{{\tr}_{+}(\jslash\, \aslash\, \lslash_2 \, \bslash)}
\def\ialtb{{\tr}_{+}(\islash\, \aslash\, \lslash_2 \, \bslash)}
\def\bltloa{{\tr}_{+}(\bslash\, \lslash_2\, \lslash_1 \, \aslash)}
\def\jbloa{{\tr}_{+}(\jslash\, \bslash\, \lslash_1 \, \aslash)}
\def\ibPb{{\tr}_{+}(\islash\, \bslash\, \Pslash \, \bslash)}
\def\ijltb{{\tr}_{+}(\islash\, \jslash\, \lslash_2 \, \bslash)}

\def\ijloa{{\tr}_{+}(\islash\, \jslash\,  \lslash_1 \, \aslash)}
\def\ijblt{{\tr}_{+}(\islash\, \jslash\,  \bslash \, \lslash_2)}

\def\jakb{{\tr}_{+}(\jslash\, \aslash\, \kslash \, \bslash)}
\def\iakb{{\tr}_{+}(\islash\, \aslash\, \kslash \, \bslash)}

\def\tofo{{\tr}_{+}(\onslash\, \thslash\, \twslash \, \foslash)}
\def\foto{{\tr}_{+}(\onslash\, \thslash\, \foslash \, \twslash)}
\def\tofi{{\tr}_{+}(\onslash\, \thslash\, \twslash \, \fislash)}
\def\fito{{\tr}_{+}(\onslash\, \thslash\, \fislash \, \twslash)}

\def\lrangle#1#2{\langle #1\,#2\rangle}

\def\Li{{$\rm Li}_2$}
\def\eps{\epsilon}
\def\epsuv{{\epsilon_{\rm \mbox{\tiny UV}}}}
\let\bm=\bibitem
\let\la=\label

\def\npb#1#2#3{Nucl. Phys. {\bf{B#1}} #3 (#2)}
\def\plb#1#2#3{Phys. Lett. {\bf{#1B}} #3 (#2)}
\def\prl#1#2#3{Phys. Rev. Lett. {\bf{#1}} #3 (#2)}
\def\prd#1#2#3{Phys. Rev. {D \bf{#1}} #3 (#2)}
\def\cmp#1#2#3{Comm. Math. Phys. {\bf{#1}} #3 (#2)}
\def\cqg#1#2#3{Class. Quantum Grav. {\bf{#1}} #3 (#2)}
\def\nppsa#1#2#3{Nucl. Phys. B (Proc. Suppl.) {\bf{#1A}}#3 (#2)}
\def\ap#1#2#3{Ann. of Phys. {\bf{#1}} #3 (#2)}
\def\ijmp#1#2#3{Int. J. Mod. Phys. {\bf{A#1}} #3 (#2)}
\def\rmp#1#2#3{Rev. Mod. Phys. {\bf{#1}} #3 (#2)}
\def\mpla#1#2#3{Mod. Phys. Lett. {\bf A#1} #3 (#2)}
\def\jhep#1#2#3{J. High Energy Phys. {\bf #1} #3 (#2)}
\def\atmp#1#2#3{Adv. Theor. Math. Phys. {\bf #1} #3 (#2)}
\newcommand{\EQ}[1]{\begin{equation} #1 \end{equation}}
\newcommand{\AL}[1]{\begin{subequations}\begin{align} #1 \end{align}\end{subequations}}
\newcommand{\SP}[1]{\begin{equation}\begin{split} #1 \end{split}\end{equation}}
\newcommand{\ALAT}[2]{\begin{subequations}\begin{alignat}{#1} #2 \end{alignat}
                        \end{subequations}}
\def\beqa{\begin{eqnarray}}
\def\eeqa{\end{eqnarray}}
\def\beq{\begin{equation}}
\def\eeq{\end{equation}}
\def\sst{\scriptscriptstyle}
\def\thetabar{\bar\theta}
\def\Tr{{\rm Tr}}
\def\one{\mbox{1 \kern-.59em {\rm l}}}
 \def\Nh{\hat{N}}

\newcommand{\half}{{\textstyle {1 \over 2}}}

\def\a{\alpha}      \def\da{{\dot\alpha}}
\def\b{\beta}       \def\db{{\dot\beta}}
\def\c{\gamma}  \def\G{\Gamma}  \def\cdt{\dot\gamma}
\def\d{\delta}  \def\D{\Delta}  \def\ddt{\dot\delta}
\def\e{\epsilon}        \def\vare{\varepsilon}
\def\f{\phi}    \def\F{\Phi}    \def\vvf{\f}
\def\h{\eta}
\def\k{\kappa}
\def\l{\lambda} \def\L{\Lambda}
\def\m{\mu} \def\n{\nu}
\def\o{\omega}
\def\p{\pi} \def\P{\Pi}
\def\r{\rho}
\def\s{\sigma}  \def\S{\Sigma}
\def\t{\tau}
\def\th{\theta} \def\Th{\Theta} \def\vth{\vartheta}
\def\X{\Xeta}
\def\z{\zeta}
\def\de{\partial}

\def\cA{{\cal A}} \def\cB{{\cal B}} \def\cC{{\cal C}}
\def\cD{{\cal D}} \def\cE{{\cal E}} \def\cF{{\cal F}}
\def\cG{{\cal G}} \def\cH{{\cal H}} \def\cI{{\cal I}}
\def\cJ{{\cal J}} \def\cK{{\cal K}} \def\cL{{\cal L}}
\def\cM{{\cal M}} \def\cN{{\cal N}} \def\cO{{\cal O}}
\def\cP{{\cal P}} \def\cQ{{\cal Q}} \def\cR{{\cal R}}
\def\cS{{\cal S}} \def\cT{{\cal T}} \def\cU{{\cal U}}
\def\cV{{\cal V}} \def\cW{{\cal W}} \def\cX{{\cal X}}
\def\cY{{\cal Y}} \def\cZ{{\cal Z}}

\def\ua{\underline{\alpha}}
\def\ub{\underline{\phantom{\alpha}}\!\!\!\beta}
\def\uc{\underline{\phantom{\alpha}}\!\!\!\gamma}
\def\um{\underline{\mu}}
\def\ud{\underline\delta}
\def\ue{\underline\epsilon}
\def\una{\underline a}\def\unA{\underline A}
\def\unb{\underline b}\def\unB{\underline B}
\def\unc{\underline c}\def\unC{\underline C}
\def\und{\underline d}\def\unD{\underline D}
\def\une{\underline e}\def\unE{\underline E}
\def\unf{\underline{\phantom{e}}\!\!\!\! f}\def\unF{\underline F}
\def\unm{\underline m}\def\unM{\underline M}
\def\unn{\underline n}\def\unN{\underline N}
\def\unp{\underline{\phantom{a}}\!\!\! p}\def\unP{\underline P}
\def\unq{\underline{\phantom{a}}\!\!\! q}
\def\unQ{\underline{\phantom{A}}\!\!\!\! Q}
\def\unH{\underline{H}}

\def\As {{A \hspace{-6.4pt} \slash}\;}
\def\bs {{b \hspace{-6.4pt} \slash}\;}
\def\Ds {{D \hspace{-6.4pt} \slash}\;}
\def\ds {{\del \hspace{-6.4pt} \slash}\;}
\def\ss {{\s \hspace{-6.4pt} \slash}\;}
\def\ks {{ k \hspace{-6.4pt} \slash}\;}
\def\ps {{p \hspace{-6.4pt} \slash}\;}
\def\pas {{{p_1} \hspace{-6.4pt} \slash}\;}
\def\pbs {{{p_2} \hspace{-6.4pt} \slash}\;}
\def\Ps {{P \hspace{-6.4pt} \slash}\;}
\def\Qs {{Q \hspace{-6.4pt} \slash}\;}

\def\Fh{\hat{F}}
\def\Vh{\hat{V}}
\def\Xh{\hat{X}}
\def\ah{\hat{a}}
\def\xh{\hat{x}}
\def\yh{\hat{y}}
\def\ph{\hat{p}}
\def\xih{\hat{\xi}}
\def\psit{\tilde{\psi}}
\def\Psit{\tilde{\Psi}}
\def\tht{\tilde{\th}}
\def\lt{\tilde{\lambda}}
\def\hl{\hat{\lambda}}
\def\hlt{\hat{\tilde{\lambda}}}
\def\llt{\tilde{l}}
\def\At{\tilde{A}}
\def\Qt{\tilde{Q}}
\def\Rt{\tilde{R}}
\def\Nt{\tilde{N}}

\def\at{\tilde{a}}
\def\st{\tilde{s}}
\def\ft{\tilde{f}}
\def\pt{\tilde{p}}
\def\qt{\tilde{q}}
\def\vt{\tilde{v}}
\def\nt{\tilde{n}}

\def\delb{\bar{\partial}}
\def\bz{\bar{z}}
\def\bD{\bar{D}}
\def\bB{\bar{B}}

\def\bk{{\bf k}}
\def\bl{{\bf l}}
\def\bp{{\bf p}}
\def\bq{{\bf q}}
\def\br{{\bf r}}
\def\bx{{\bf x}}
\def\by{{\bf y}}
\def\bR{{\bf R}}
\def\bV{{\bf V}}

\def\d{\delta}\def\D{\Delta}\def\ddt{\dot\delta}
\def\pa{\partial} \def\del{\partial}
\def\xx{\times}
\def\uno{\mbox{1 \kern-.59em {\rm l}}}
\def\trp{^{\top}}
\def\inv{^{-1}}
\def\dag{{^{\dagger}}}
\def\pr{^{\prime}}
\def\lan{\langle}
\def\ran{\rangle}
\def\rar{\rightarrow}
\def\lar{\leftarrow}
\def\lrar{\leftrightarrow}
\newcommand{\0}{\,\!}      
\def\one{1\!\!1\,\,}
\def\im{\imath}
\def\jm{\jmath}
\newcommand{\tr}{\mbox{tr}}
\newcommand{\slsh}[1]{/ \!\!\!\! #1}
\def\vac{|0\rangle}
\def\lvac{\langle 0|}
\def\hlf{\frac{1}{2}}
\def\ove#1{\frac{1}{#1}}
\def\Box{\square}
\def\ZZ{\mathbb{Z}}
\def\CC#1{({\bf #1})}
\def\bcomment#1{}
\def\bfhat#1{{\bf \hat{#1}}}
\def\VEV#1{\left\langle #1\right\rangle}
\newcommand{\ex}[1]{{\rm e}^{#1}} \def\ii{{\rm i}}
\def\rr{{\rm r}} \def\rs{{\rm s}}\def\rv{{\rm v}}
\def\ri{{\rm i}}\def\rj{{\rm j}}
\newcommand{\lrbrk}[1]{\left(#1\right)}
\newcommand{\sfrac}[2]{{\textstyle\frac{#1}{#2}}}

\def\Li{{\rm Li}_2}

\font\mybb=msbm10 at 12pt
\def\bb#1{\hbox{\mybb#1}}

\font\myBB=msbm10 at 18pt
\def\BB#1{\hbox{\myBB#1}}

\begin{document}

\begin{flushright}
QMUL-PH-11-08\\
NSF-KITP-11-146
\end{flushright}

\vspace{20pt}

\begin{center}

{\Large \bf    Harmony of Super Form Factors   }
\vspace{11pt}
\vspace{32pt}

{\mbox {\bf A.~Brandhuber$^{a,b}$,  \"O.~G\"urdo\u gan$^{a}$,  R.~Mooney$^{a}$, G.~Travaglini$^{a,b}$ and G.~Yang$^{a}$}}%
\footnote{ {\sffamily \{\tt a.brandhuber,o.c.gurdogan, r.j.b.mooney,  g.travaglini,
g.yang\}@qmul.ac.uk }}

\bigskip

$^{a}${\em Centre for Research in String Theory\\
School of Physics and Astronomy\\
Queen Mary University of London\\
Mile End Road, London, E1 4NS\\
United Kingdom

\vskip .3truecm
$^{b}$Kavli Institute for Theoretical Physics\\ University of California, Santa Barbara, CA 93106, USA
 }

\bigskip

\vspace{30pt} {\bf Abstract}

\end{center}

\noindent
In this paper we continue our systematic study of form factors of half-BPS operators in $\cN\!=\!4$ super Yang-Mills.
In particular,  we extend various on-shell techniques known for amplitudes to the case of form factors, including MHV rules,
recursion relations, unitarity and dual MHV rules.  As an application, we present the solution of the recursion relation for split-helicity form factors.
We then consider form factors of the stress-tensor multiplet operator and of its chiral truncation, and write down supersymmetric Ward identities using chiral as well as non-chiral superspace formalisms. This allows us to obtain compact formulae for families of form factors, such as the maximally non-MHV case.
Finally we generalise dual MHV rules in dual momentum space to form factors.

\setcounter{page}{0}
\thispagestyle{empty}
\newpage


\setcounter{tocdepth}{4}
\hrule height 0.75pt
\tableofcontents
\vspace{0.8cm}
\hrule height 0.75pt
\vspace{1cm}

\setcounter{tocdepth}{2}


\setcounter{footnote}{0}

\section{Introduction}

Form factors are interesting physical observables which are situated at the interface between completely on-shell quantities such as scattering amplitudes and completely off-shell quantities like correlation functions.
In a gauge theory one typically considers the overlap of a state created by a gauge-invariant operator $\cO(x)$ with a multiparticle state $\lan 1 \cdots n |$ described by the particles' momenta $p_1, \ldots , p_n$ and other relevant quantum numbers such as the helicity, for massless particles. We will usually consider the Fourier transform of the form factor,
\beq
\int\!\!d^4x \, e^{-iqx} \, \lan 1 \cdots n | \cO (x) |0\ran \ =   \ \delta^{(4)} (q - \sum_{i=1}^n p_i ) \, \lan 1 \cdots n | \cO (0) |0\ran \\
\ ,
\eeq
where the momentum delta function appears as a consequence of translational invariance of the theory,  with $\cO(x) = \exp (i \cP x ) \cO(0) \exp ( - i \cP x ) $.

Form factors appear in several interesting physical contexts.  Some of the early applications include the amplitude for deep inelastic scattering,
which is controlled by  the matrix element $\lan X | J^{\rm e.m.}_{{\rm h, }\mu}(0)|p\ran$ of the hadronic electromagnetic current
$J^{\rm e.m.}_{{\rm h, }\mu}$ with an initial proton state $p$ and a final hadronic state $X$, and the
$e^+ e^-\to X$ annihilation process,  governed by the form factor   $\lan X | J^{\rm e.m.}_{{\rm h, }\mu}(0)|0\ran$.
Furthermore, the universal structure of infrared divergences of amplitudes is controlled by the
Sudakov form factor   \cite{ir1, ir2, ir3, ir4, ir5,  ir6, ir7}, with the coefficient of the leading infrared divergence being related to the cusp anomalous dimension \cite{ir-wl1} and the large-spin limit of  twist-two operators \cite{ir-wl2}. Its universal, exponential form
has inspired the all-loop conjecture for planar MHV amplitudes in $\cN=4$ super Yang Mills (SYM) \cite{bds}. Another interesting application of form factors is the operator product expansion of null polygonal Wilson loops proposed recently in \cite{wl-ope}.

Ultimately, one of the important goals and motivations for the study of form factors in $\cN=4$ SYM is that they interpolate between off-shell and on-shell quantities. For off-shell quantities such as two-point correlation functions, integrability has been developed into a powerful computational tool,%
\footnote{For a recent review see \cite{Beisert:2010jr}  and references therein. }
and it also plays an important role in the calculation of amplitudes \cite{Alday:2009dv,Alday:2010vh}
and form factors \cite{Alday:2007he, mz} at strong coupling.
On the other hand, for amplitudes at weak coupling we have so far only glimpses of hidden integrable structures, and we hope that this interpolation
will give us new insights on the role and uses of integrability in the context of  amplitudes.

Form factors in maximally $\cN\!=\!4$ super Yang-Mills at strong coupling were recently considered in \cite{Alday:2007he, mz}. At weak coupling, they  were first considered in  \cite{VN1}, and in greater detail and generality in \cite{bsty2} and \cite{Bork:2010wf}.
In particular, \cite{bsty2} considered form factors of the half-BPS scalar operator $\Tr (\phi_{12} \phi_{12})$ in $\cN\!=\!4$ SYM at tree level and one loop, where $\phi_{AB}$ are the six scalar fields in the theory, $A,B=1, \ldots , 4$, with $\phi_{AB}= -\phi_{BA}$,  with external states containing two scalars and an arbitrary number of positive-helicity gluons. These MHV form factors were found to be remarkably simple. Specifically,
at tree level  they are expressed in terms of a holomorphic function of the spinor variables associated to the particle momenta which is a close cousin of the Parke-Taylor MHV scattering amplitude \cite{Mangano:1987xk}. This simplicity was also found to persist at one loop, where the  result for these MHV form factors is a remarkably simple expression which is  very reminiscent of that for an $n$-point MHV amplitude at one loop.

In this paper we continue the systematic study initiated in  \cite{bsty2}.  The preceding discussion  has already outlined two of the motivations for this study. Firstly, the attempt at connecting the on-shell world of scattering amplitudes (with a flurry of new techniques discovered over the past seven years) to that of off-shell observables in the theory, with integrability playing a prominent role in determining the quantum structure of some of these observables at strong and weak coupling.
 The second motivation is that form factors are expressed by simple  formulae  despite being partially off shell. As we shall see, certain form factors exhibit further unexpected simplicities, such as the maximally non-MHV form factors.  More concretely, in this paper we pursue the following objectives.

In Section 2 we begin by extending to form factors some distinguished  on-shell techniques used successfully over the past years to calculate amplitudes, such as MHV diagrams \cite{csw} and on-shell recursion relations \cite{bcf, bcfw}.
Some of these findings are not unexpected -- for example, MHV rules are related to an MHV  Lagrangian which is applicable also off shell
 \cite{Mansfield:2005yd}, and recursion relations are based on factorisation, which is a general property not only of amplitudes but also of  Green's functions
 \cite{wein}. Notice that MHV diagrams are expected to work also in the presence of multiple operator insertions.
As an application, we will explicitly solve the  recursion relations for form factors where the external state is made of gluons in a split-helicity configuration. This parallels a corresponding explicit solution for amplitudes found in \cite{BFRSV06}.

Section 3 --- a central part of this paper --- is devoted to supersymmetric form factors. We will find that  harmonic superspace
\cite{harm1, harm2}  is a very convenient framework to formulate and study such objects. More precisely, we consider form factors  where the operator inserted is the {\it chiral part} of the stress-tensor multiplet operator, which preserves half of the supersymmetries {\it off shell} \cite{Eden:2011yp,Eden:2011ku}, while the state is described using the supersymmetric formalism of Nair \cite{Nair}.  Using a chiral superspace formulation
we write down supersymmetric Ward identities for these form factors, and show how they constrain their form.
As a particular application of these techniques we derive the maximally non-MHV form factors which turn out to be surprisingly simple. Finally, we study form factors of the full stress-tensor multiplet operator for which we introduce  a non-chiral superspace representation.

In Section 4 we briefly introduce supersymmetric MHV diagrams, supersymmetric recursion relations
and unitarity for  form factors. These generalisations are rather straightforward, so our presentation here is somewhat condensed.

Finally, in Section 5 we introduce dual MHV rules  for form factors
formulated directly in dual (super) momentum space
by giving various examples at tree and one-loop level, and point out and explain certain subtleties encountered at higher loops. In this construction, a certain  periodic kinematic configuration, emerging in the strong-coupling calculation of \cite{Alday:2007he, mz},  plays a central role,
and we discuss similarities with (and differences from) the Wilson loop/amplitude duality \cite{am, dks,bht}.

The understanding of the large-$z$ behaviour of form factors is crucial in formulating the recursion relation, and is analysed in detail in Appendix A. Finally, in Appendix B we provide a brief reminder of the dual MHV rules for  amplitudes.


\section{Tree-level methods}

In this section we will develop and extend tree-level methods for form factors by generalising
the corresponding methods for amplitudes, namely MHV diagrams \cite{csw} and on-shell
recursions relations \cite{bcf, bcfw}.%
\footnote{For a recent review of tree-level methods in gauge theory and gravity, see \cite{bstrev}.}
We then proceed to obtain several new results including the NMHV and all split-helicity cases. We will not present the calculations with both methods for all examples but wish to stress here that we have made extensive checks to confirm that the results obtained with either method always agree. The supersymmetrisation of these methods will be considered in Section 4.

\subsection{MHV diagrams \label{section-boson}}

We start with a simple extension of the MHV diagram method \cite{csw} to form factors.
We will test this here only in  tree-level calculations, but the extension to loop level,
following \cite{bst}, is straightforward.

Specifically, we will be interested in calculating NMHV form factors of the simplest class of operators in $\cN=4$ SYM, namely the half-BPS operators  $ \Tr (\phi_{12} \phi_{12} )$.
They take the form
\beq
\lan \, g^+ (p_1)  \cdots  \phi_{12} (p_i) \cdots   \phi_{12} (p_j) \cdots   g^+(p_{n-1})\,   g^-  (p_n) \,  | \Tr (\phi_{12} \phi_{12} )(x) | \, 0 \, \ran\ ,
\eeq
where all but one of the gluons have positive helicity.
The strategy of the calculation is very simple --  we  need to augment the set of usual MHV vertices for amplitudes by including
 a new family of MHV vertices, obtained by continuing off shell  the tree-level MHV form factors  of the half-BPS operators.
 The expressions for these quantities were derived in \cite{bsty2}, and are given by
\beqa
&&\int\!\!d^4x\, e^{-iqx} \, \lan  g^+(p_1) \cdots  \phi_{12} (p_i) \cdots
 \phi_{12} (p_j) \cdots g^+ (p_n)   |  \Tr (\phi_{12} \phi_{12} )(x) | 0 \ran
 \nonumber \\
 &&=
 g^{n-2} (2 \pi)^4 \delta^{(4)} ( \sum_{k=1}^n \l_k \lt_k - q) \ F_{\rm MHV}
\ ,
\eeqa
where
\beq
\label{ffmhvtree}
F_{\rm MHV}\ = \   {\lan ij\ran^2 \over \lan 12\ran \cdots \lan n1\ran }\ .
\eeq
Here  $p_m := \l_m \lt_m$ are on-shell momenta of the external particles, and $q := \sum_{m=1}^n p_m$ is the momentum carried by the operator insertion.
It was observed in  \cite{bsty2} that, since \eqref{ffmhvtree}  is a holomorphic function of the spinor variables, the MHV form factors are  localised on a complex line in twistor space, similarly to the MHV amplitudes \cite{witten}.

Using localisation as an inspiration, we propose to use an appropriate  off-shell continuation of \eqref{ffmhvtree} as a new vertex
to construct the perturbative expansion of non-MHV form factors of the operator $ \Tr (\phi_{12} \phi_{12} )$. The off-shell continuation  is the standard one introduced in \cite{csw}. The momentum $L$ of an internal, off-shell particle is decomposed as
$L  \ = \ l + z \xi$, where  $l = \l_L \lt_L$ is an on-shell momentum and  $\xi$ an arbitrary  reference null momentum.
The off-shell continuation of  \cite{csw} consists then in using the spinor $\l_L$ as the spinor variable associated with the internal leg of momentum $L$, where
\beqa
\label{off1}
\l_{L, \a} & = & {L_{\a \da} \tilde{\xi}^{\da}
\over [ \lt_L \,  , \tilde{\xi}]
}
\ .
 \eeqa
The denominator in the right-hand side of \eqref{off1}
will be irrelevant for our applications since each MHV diagram is invariant under rescalings of the internal spinor variables.  Hence, we will discard it and simply replace $\l_{L, \a} \to  L_{\a \da} \tilde{\xi}^{\da}$.

\subsubsection{NMHV form factors}

Using the MHV rules outlined in the previous section, we now present
an example of derivation of an NMHV form factor. Specifically,
the form factor we consider is
\beq
 \label{1stffnmhv}
 F_{\rm NMHV}
(1_{\phi_{12}} , 2_{\phi_{12}}, 3_{g^-}, 4_{g^+})  := \lan \phi_{12}
(p_1)\phi_{12} (p_2) g^-(p_3) g^+ (p_4) | \Tr (\phi_{12} \phi_{12}
)(0) |  0 \ran  \ .
\eeq
%
%
\begin{figure}[h]
\centerline{\includegraphics[height=5cm]{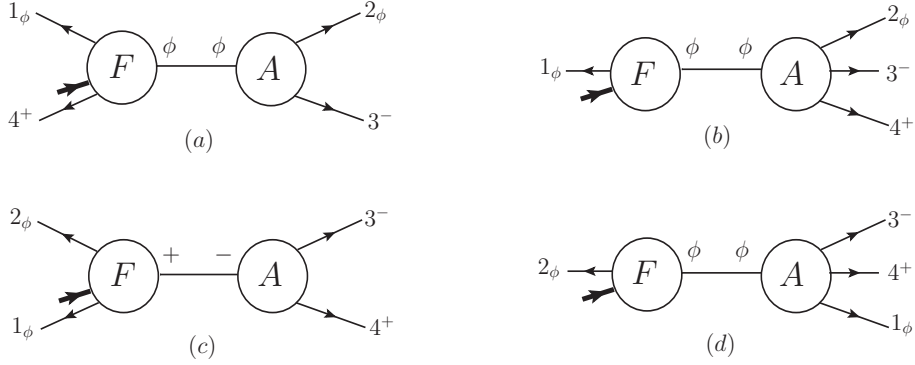} } \caption{\it The four MHV diagrams contributing to the NMHV
form factor \eqref{1stffnmhv}.} \label{MHVfig}
\end{figure}
%
%
There are four MHV diagrams contributing to
\eqref{1stffnmhv}, depicted in Figure \ref{MHVfig}. A
short calculation shows that these are given by the following
expressions:
\beqa {\rm Diagram} \ (a) & = & {[2 \xi]\over [\xi 3]
} { 1\over [32] \lan 41 \ran} { \lan 1 | q-p_4 | \xi] \over |\lan 4
| q-p_1 | \xi] } \ ,
\nonumber \\
{\rm Diagram} \ (b) & = &
{\lan 23 \ran \over \lan 34 \ran s_{234} } { \lan 3 | p_{2}  + p_4 | \xi]^2 \over\lan 2 |  p_3 + p_4 | \xi ] \lan 4| p_2 + p_3  | \xi]}
\ ,
\nonumber \\
{\rm Diagram} \ (c) & = & {\lan 12 \ran \over [43]} { [\xi 4]^3 \over [3 \xi]} {1 \over \lan 2 | p_3 + p_4 | \xi] \lan 1 | p_3 + p_4 | \xi]}
\ ,
\nonumber \\
{\rm Diagram} \ (d) & = &
{1\over s_{341}} {\lan 13\ran^2 \over \lan 34\ran \lan41 \ran} { \lan 3| p_4 + p_1 | \xi] \over \lan 1 | p_3 + p_4 | \xi]}
\ .
 \eeqa
We have checked that the sum of all MHV diagrams is independent of the choice of the reference spinor $\tilde\xi$. A particularly convenient choice of $\tilde\xi$  is $\tilde\xi = \tilde\lambda_4$,  in which case we get
 \beqa
 \label{MHVfn}
 F_{\rm NMHV} (1_{\phi_{12}} , 2_{\phi_{12}}, 3_{g^-}, 4_{g^+}) & = &
{[24]\over [34] } {1\over \lan 4 | p_2 + p_3 | 4 ] } \Big[ {\lan 1 | q | 4] \over [23] \lan 41 \ran } + { [24] \lan 23 \ran^2 \over \lan 34 \ran} {1 \over s_{234}} \Big]
\nonumber \\
 &+&
 {\lan 13 \ran^2 [14] \over \lan 41 \ran \lan 34 \ran [43]} {1\over s_{341}}
\ .
\eeqa
It is straightforward to apply this procedure to more general form factors but for
brevity we will not present them here. However, we mention that all results derived in the next subsection using recursion relations have been compared with formulae obtained from MHV diagrams finding a perfect match in all cases.


\subsection{Recursion relations}

In this subsection we study the application of recursion relations to the derivation of tree-level form factors. As a warm-up we will re-derive the NMHV form factor in \eqref{1stffnmhv} finding agreement with \eqref{MHVfn}, and then move on to consider more general cases including split-helicity configurations. Since form factors contain a single operator insertion, it is clear that every recursive diagram will contain one amplitude and one form factor as the factorisation properties used in the case of tree-level recursions for amplitudes also apply to tree-level form factors. This is the only modification to
the on-shell recursion relations of \cite{bcf}. In Appendix A we discuss the behaviour of form factors under large complex deformations, and confirm the validity of the calculations below, i.e.~we show that under the shifts used the form factors vanish as $z \to \infty$.

Let us begin by re-deriving the NMHV form factor  \eqref{1stffnmhv}. We will use a $[34\ran$ shift, namely
\beq
\hat{\tilde{\lambda}}_3 := \tilde{\lambda}_3 + z \tilde{\lambda}_4 \ , \qquad \qquad \hat\lambda_4 := \lambda_4 - z \lambda_3
\ .
\eeq
There are two recursive diagrams, depicted in Figure \ref{BCFWfig} below. A short calculation shows that
\beqa
{\rm Diagram} \ (a) & = &  {[24]^2 \over [23][34]}{1\over s_{234}} {\lan 1 | q | 4 ] \over \lan 1 | q | 2 ] }
\ ,
\nonumber \\
{\rm Diagram} \ (b) &=& {\lan 13\ran^2 \over \lan 34 \ran \lan 41 \ran} {1\over s_{341}} {\lan 3 | q | 2 ] \over \lan 1 | q | 2 ] }
\ ,
\eeqa
so that
\beq
 F_{\rm NMHV} (1_{\phi_{12}} , 2_{\phi_{12}}, 3_{g^-}, 4_{g^+}) =  {1\over \lan 1 | q | 2 ]}\left[
{[24]^2 \over [23][34]}{1\over s_{234}} {\lan 1 | q | 4 ]  }  \ + \   {\lan 13\ran^2 \over \lan 34 \ran \lan 41 \ran} {1\over s_{341}} {\lan 3 | q | 2 ]}\right]\,.
\eeq
It is interesting to note that the $1/\lan 1 | q | 2 ]$ pole is in fact spurious. This can be shown by using the identities
\beqa
\lan 1 | q \, p_4 | 3\ran + \lan 1 | q\,  p_2 | 3 \ran &= &\lan 13\ran s_{234}
\ ,
\nonumber \\
{[4 | p_3\,  q | 2 ]} + { [4 | p_1 \, q | 2 ] } &=& [42] s_{341}
\ ,
\eeqa
which allow to recast the form factor in the alternative form
\beq
\label{recfn}
 F_{\rm NMHV} (1_{\phi_{12}} , 2_{\phi_{12}}, 3_{g^-}, 4_{g^+}) =  {1\over s_{34}  \, [23] \lan41 \ran }\left[ { \lan 14\ran \lan 23\ran [24]^2 \over s_{234}} + { [41] [32] \lan 13\ran^2 \over s_{341 }} + [24] \lan 13\ran \right]\, .
\eeq
We have checked that our result \eqref{MHVfn} for the form factor derived using  MHV diagrams, and  \eqref{recfn}, obtained using recursion relations, are in agreement.
%
%
\begin{figure}[h]
\centerline{\includegraphics[height=2.5cm]{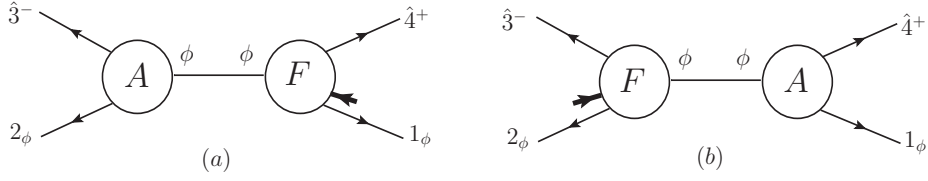} } \caption{\it
The two recursive diagrams contributing
to the NMHV form factor \eqref{1stffnmhv}.}
\label{BCFWfig}
\end{figure}
%
%


\subsubsection{Recursion relations for the split-helicity  form factor \label{section-splithelicity}}

\newcommand{\ang}[1]
{
  \langle #1 \rangle
}

\newcommand{\sqr}[1]
{
  [ #1 ]
}
\newcommand{\tlambda}
{
  \tilde{\lambda}
}

\newcommand{\MHVbar}
{
  $\overline{\rm MHV}$
}
In the previous section we found that the BCF recursion relation for the NMHV form factor with a $[3,4 \rangle$ shift
has just two diagrams.  This property in fact holds for all form factors of the form $F_{\phi^2;q-2,n-q}(1_\phi,2_\phi,3^-,\ldots,q^-,(q+1)^+,\ldots,n^+)$, which we call henceforth \emph{split-helicity}. As we will show shortly, performing a  $[q,q+1\rangle$ shift leads to a general, closed-form solution of the BCFW recursion relations for this special class of form factors. Note that all split-helicity gluon scattering amplitudes were computed in \cite{BFRSV06} -- we construct here a similar solution for form factors.

Each recursive diagram with a $[q,q+1\rangle$ shift contains  a three-point amplitude and an $(n-1)$-point form factor. We can neatly combine the three-point amplitude and the propagator in a prefactor to write%
\footnote{For the rest of this section we will always assume that the operator $\cO=\Tr(\phi_{12}\phi_{12})$ is inserted and will not mention it explicitly. Although the solution is presented for this particular insertion, the construction can be generalised to form factors involving other operators.}
\begin{equation}
\label{eq:splitrec}
F_{q-2,n-q}\begin{aligned}[t]&=\frac{[q-1 q+1]}{[q-1 q][q q+1]}F_{q-3,n-q}(1_\phi,2_\phi,3^-,\ldots,\widehat{q-1}^-,\widehat{q+1}^+,\ldots,n^+)\\
&+ \frac{\lan q  q+2 \ran}{\lan q q+1 \ran \lan q+1 q+2 \ran}F_{q-2,n-q-1}(1_\phi,2_\phi,3^-,\ldots,\hat{q}^-,\widehat{q+2}^+,\ldots n^+)
\, ,
\end{aligned}
\end{equation}
where the shifted spinors of the external momenta that appear in the lower-point form factors  are
\begin{subequations}
  \begin{align}
    \lambda_{\widehat{q+1}} &=\frac{[q-1| P_{q,q+1}}{\sqr{q-1\,q+1}}\, , \\
    \tlambda_{\widehat{q}} &=\frac{P_{q,q+2}|q+2\rangle}{\ang{q\,q+2}}\, ,
  \end{align}
\end{subequations}
with $P_{a,b} = p_a + \ldots + p_b$. Furthermore,
the shifted spinors associated with internal legs are relabelled as
\begin{subequations}
  \begin{align}
    \lambda_{\hat{P}_{q-1\,q}}\left(z=z_{q-1\,q}\right)\ \rightarrow \ \lambda_{\widehat{q-1}} & \ = \ \frac{P_{q,q+1} | q+1]}{\sqr{q-1\, q+1}}
    \, ,
    \\
    \tlambda_{\hat{P}_{q+1\,q+2}}\left(z=z_{q+1\,q+2}\right)\ \rightarrow \ \tlambda_{\widehat{q+2}} & \ = \ \frac{\langle q | P_{q,q+2}}{\ang{q\,q+2}}\, ,
  \end{align}
\end{subequations}
so that the notation remains compatible with subsequent recursions.
Crucially, all lower-point form factors appearing in \eqref{eq:splitrec} are of split-helicity form, so that the split helicity form factors are closed under  recursions. Once we have reduced the form factor to expressions that involve only MHV and $\overline{\rm MHV}$ terms, we can insert the shifted momenta.

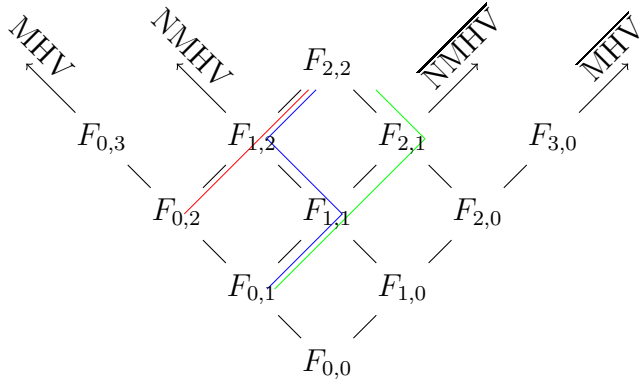
\begin{figure}[h]
\centering
\begin{tikzpicture}[on grid]
	
	\draw[red ,line width=0.3] (-0.9,3) -- (-1.9,2);
	\draw[blue, line width=0.3] (-0.8,3) -- (0.2,2) -- (-0.8,1);
	\draw[green, line width=0.3] (1.3,3) -- (0.3,2) -- (-0.7,1);

	\node (F00) 				{$F_{0,0}$};

	\node (F10) [above right=of F00] 	{$F_{1,0}$};
	\node (F01) [above left=of F00] 	{$F_{0,1}$};

	\draw (F00) -- (F10);
	\draw (F00) -- (F01);

	\node[dotted] (F20) [above right=of F10] 	{$F_{2,0}$};
	\node[dotted] (F11) [above left=of F10] 	{$F_{1,1}$};
	\node[dotted] (F02) [above left=of F01] 	{$F_{0,2}$};

	\draw (F01) -- (F02);
	\draw (F01) -- (F11);
	\draw (F10) -- (F11);
	\draw (F10) -- (F20);

	\node (F30) [above right=of F20] 	{$F_{3,0}$};
	\node (F21) [above left =of F20] 	{$F_{2,1}$};
	\node (F12) [above right=of F02] 	{$F_{1,2}$};
	\node (F03) [above left=of F02] 	{$F_{0,3}$};

	\draw (F02) -- (F03);
	\draw (F02) -- (F12);
	\draw (F11) -- (F12);
	\draw (F11) -- (F21);
	\draw (F20) -- (F21);
	\draw (F20) -- (F30);

	\node (F22) [above left=of F21]		{$F_{2,2}$};
	\node (F221) at (0.1,4)	{\phantom{$F_{2,2}$}};
	\node (F222) at (0.2,4)	{\phantom{$F_{2,2}$}};
	\node (F223) at (0.3,4) {\phantom{$F_{2,2}$}};

	\draw (F12) -- (F22);
	\draw (F21) -- (F22);
	\draw[->] (F21) -- +(1,1);
	\draw[->] (F12) -- +(-1,1);

	\draw[red,line width=0.3] (F221) to (-0.9,3);
	\draw[blue,line width=0.3] (F222)  to (-0.8,3);
	\draw[green, line width=0.3] (F223) to (1.3,3);
	
	\draw[->]	(F03) -- +(-1,1) node[above,rotate=-45]{MHV};
	\draw[->]	(F30) -- +(1,1)  node[above,rotate=45]{$\overline{\rm{MHV}}$};
	\draw		(-2,4) node[above,rotate=-45]{NMHV};
	\draw		(2,4) node[above,rotate=45]{$\overline{\rm{NMHV}}$};

\end{tikzpicture}
\caption{\it The iterative structure of split-helicity form factors illustrated by a square lattice. The three coloured  paths ending on the MHV line are in one-to-one correspondence with terms that appear in the iterated recursion of $F_{2,2}$. Similarly there will be three paths (terms) that end on the \MHVbar line.}
\label{lattice}
\end{figure}

It is useful to illustrate the structure of the recursion relations for split-helicity form factors using
a square lattice as in Figure \ref{lattice}.
Consider for example the form factor $F_{2,2}$. In this case,  the first iteration using equation (\ref{eq:splitrec}) relates $F_{2,2}$ to the form factors $F_{2,1}$ and $F_{1,2}$, which however are neither $\rm MHV$ nor $\overline{\rm MHV}$. The next iteration leads to an expression involving one $F_{2,0}$, two $F_{1,1}$'s and one $F_{0,2}$ evaluated at some shifted momenta.
A final iteration would then allow us to express the answer in terms of
$\rm MHV$ and $\overline{\rm MHV}$ form factors alone, or even to reduce everything down to $F_{0,0}$.
It is also easy to see that this pattern generalises to arbitrary split-helicity form factors and that each term generated by subsequent recursions corresponds to a unique path between the form factor and the $\rm MHV$ or $\overline{\rm MHV}$ edges of the lattice, as illustrated in Figure \ref{lattice}.

In principle, all we need to do to compute a split-helicity form factor is to collect all prefactors picked up at each step of the recursion process and follow the iterated momentum shifts along a particular path on the lattice.

\subsubsection{Solution for the split-helicity form factor}
A very efficient way to organise the recursion is in terms of \emph{zig-zag diagrams}, like those introduced in \cite{BFRSV06} for split-helicity gluon amplitudes. It is natural to split the terms of the solution into those corresponding to paths ending on the MHV or $\overline{\rm{MHV}}$ lines, respectively.

Zig-zag diagrams that correspond to recursion terms with an MHV form factor will be denoted as MHV zig-zags and the ones with an \MHVbar form factor as \MHVbar zig-zags. Note that we have therefore two types of diagrams, in contrast to the case of amplitudes in \cite{BFRSV06}. One can make this separation also for amplitudes as it only means that we terminate the iterated recursion once we reach an \MHVbar term, instead of recursing it further down to $F_{0,0}$ (or $A_{2,2}$ for the case of amplitudes). In the path picture of the previous section, this separation corresponds to the fact that there is a unique path between any \MHVbar form factor and $F_{0,0}$, hence one can replace that part of the recursion directly with an \MHVbar form factor. Because the MHV zig-zags defined below are not compatible with two point objects such as $F_{0,0}$ we chose to use this formalism with two types of diagrams. This has the added advantage that it makes the parity symmetry of $F_{q-2,q-2}$ form factors manifest.

The MHV zig-zags are parameterised with $2k+1$ labels
$$2\leq a_1<\dotsm < a_{k}< q-1\quad \text{and}\quad n\geq b_1 > \dotsm > b_{k+1} > q,\qquad k\geq0,$$
representing expressions in the following manner
\begin{equation}\label{eq:zzh}
 \begin{minipage}{0.65\linewidth} \begin{tikzpicture}
	\draw (0.5,1) -- (0,1) arc (90:270:1) -- +(0.5,0);
        \draw[dotted] (0.5,1) -- (1,1);
        \draw (1,1) -- (5,1);
        \draw[dotted] (5,1) -- (5.5,1);
        \draw (5.5,1) -- (7.5,1) arc (90:-90:1) -- (6,-1);
        \draw[dotted] (6,-1) -- (5.5,-1);
        \draw (5.5,-1) -- (3.5,-1);
        \draw[dotted] (3.5,-1) -- (3,-1);
        \draw (3,-1) -- (1,-1);
        \draw[dotted] (1,-1) -- (0.5,-1);
        \filldraw[fill=gray, draw=black]  (150:1) circle (3pt) node[left]{2};
        \filldraw[fill=gray, draw=black]  (210:1) circle (3pt) node[left]{1};
	\filldraw[black] 
	                        (0,-1) circle (3pt) node[below=4pt]{$n$}
				(1.5,-1) circle (3pt) node[below=4pt]{$b_1+1$}
				(2.5,-1) circle (3pt) node[below=4pt]{$b_1$}
				(4,-1) circle (3pt) node[below=4pt]{$b_2+1$}
				(5,-1) circle (3pt) node[below=4pt]{$b_2$}
				(6.5,-1) circle (3pt) node[below=4pt]{$q+2$}
				(7.5,-1) circle (3pt) node[below=4pt]{$q+1$};
	\filldraw[fill=white, draw=black]
				(2.5,1) circle (3pt) node[above=4pt]{$a_1$}
				(3.5,1) circle (3pt) node[above=4pt]{$a_1+1$}
				(6.5,1) circle (3pt) node[above=4pt]{$q-1$}
				(7.5,1) circle (3pt) node[above=4pt]{$q$};
	\draw (-1,0) .. controls (0,0) .. (2,-1) -- (3, 1) -- (4.5,-1);
	\draw         	(4.5,-1) -- +(0.25,0.5);
	\draw[dotted]	(4.75,-0.5) -- +(0.25,0.5);
	\draw      	(7,1)    -- +(-0.25,-0.5);
	\draw[dotted]	(6.75,0.5)    -- +(-0.25,-0.5);
\end{tikzpicture}\end{minipage}  =\frac{N_1N_2N_3}{D_1D_2D_3}
\end{equation}
while the \MHVbar zig-zags are parametrised with $2k+1$ labels $$2\leq \bar{b}_1<\dotsm < \bar{b}_{k+1} < q\quad \text{and}\quad n\geq \bar{a}_1 > \dotsm > \bar{a}_{k} > q+1,\qquad k\geq 0,$$ representing expressions, similarly shown below
\begin{equation}\label{eq:zzah}
 \begin{minipage}{0.65\linewidth} \begin{tikzpicture}
	\draw (0.5,1) -- (0,1) arc (90:270:1) -- +(0.5,0);
        \draw[dotted] (0.5,1) -- (1,1);

        \draw (1,-1) -- (5,-1);
        \draw[dotted] (5,-1) -- (5.5,-1);
        \draw (5.5,-1) -- (7.5,-1) arc (90:270:-1) -- (6,1);
        \draw[dotted] (6,1) -- (5.5,1);
        \draw (5.5,1) -- (3.5,1);
        \draw[dotted] (3.5,1) -- (3,1);
        \draw (3,1) -- (1,1);

        \draw[dotted] (1,-1) -- (0.5,-1);
        \filldraw[fill=gray, draw=black]  (150:1) circle (3pt) node[left]{2};
        \filldraw[fill=gray, draw=black]  (210:1) circle (3pt) node[left]{1};
	\filldraw[fill=white, draw=black] 	
                                (0,1) circle (3pt) node[above=4pt]{$3$}
				(1.5,1) circle (3pt) node[above=4pt]{$\bar{b}_1$}
				(2.5,1) circle (3pt) node[above=4pt]{$\bar{b}_1+1$}
				(4,1) circle (3pt) node[above=4pt]{$\bar{b}_2$}
				(5,1) circle (3pt) node[above=4pt]{$\bar{b}_2+1$}
				(6.5,1) circle (3pt) node[above=4pt]{$q-1$}
				(7.5,1) circle (3pt) node[above=4pt]{$q$};
	\filldraw[fill=black]
                                (0,-1) circle (3pt) node[below=4pt]{$n$}
				(2.5,-1) circle (3pt) node[below=4pt]{$\bar{a}_1+1$}
				(3.5,-1) circle (3pt) node[below=4pt]{$\bar{a}_1$}
				(6.5,-1) circle (3pt) node[below=4pt]{$q+2$}
				(7.5,-1) circle (3pt) node[below=4pt]{$q+1$};
	\draw (-1,0) .. controls (0,0) .. (2,1) -- (3, -1) -- (4.5,1);
	\draw         	(4.5,1) -- +(0.25,-0.5);
	\draw[dotted]	(4.75,0.5) -- +(0.25,-0.5);
	\draw      	(7,-1)    -- +(-0.25,0.5);
	\draw[dotted]	(6.75,-0.5)    -- +(-0.25,0.5);
\end{tikzpicture}\end{minipage}  =\frac{\bar{N_1} \bar{N_2} \bar{N_3}}{\bar{D_1} \bar{D_2} \bar{D_3}}
\end{equation}
{where $N_{1,2,3}$ and $D_{1,2,3}$ are defined as
\begin{subequations}
\begin{align}
  \begin{split}
  N_1&=\begin{aligned}[t]&\langle1|P_{2,b_1}P_{a_1+1,b_1}P_{a_1+1,b_2}P_{a_2+1,b_2}\dotsm P_{q,b_{k+1}}|q\rangle\\&\times[2|P_{a_1+1,b_1}P_{a_1+1,b_2}P_{a_2+1,b_2}\dotsm P_{q,b_{k+1}}|q\rangle^2\end{aligned}\\
  N_2&=\spaa[b_1+1\,b_1]\spaa[b_2+1\,b_2]\dotsm\spaa[b_{k+1}+1\,b_{k+1}]\\
  N_3&=[a_1 a_1+1]\dotsm[a_k\,a_k+1]\\
  D_1&=P_{2,b_1}^2P_{a_1+1,b_1}^2P_{a_1+1,b_2}^2P_{a_2+1,b_2}^2\dotsm P_{q,b_{k+1}}^2\\
  D_2&=Z_{q,1}\bar{Z}_{2,q-1}\\
  D_3&=[2|P_{2,b_1}|b_1+1\rangle\langle b_1|P_{a_1+1,b_1}|a_1][a_1+1|P_{a_1+1,b_2}|b_2+1\rangle\dotsm\langle b_{k+1}|P_{q,b_{k+1}}|q-1]
  \end{split}
\end{align}
\begin{align}
  \begin{split}
    \bar{N}_1&=\begin{aligned}[t]&[q+1|P_{\bar{b}_{k+1}+1,q+1},\dotsc,P_{\bar{b}_2+1,\bar{a}_2},P_{\bar{b}_2+1,\bar{a}_1},P_{\bar{b}_1+1,\bar{a}_1}|1\rangle^2\\&\times [q+1|P_{\bar{b}_{k+1}+1,q+1},\dotsc,P_{\bar{b}_2+1,\bar{a}_2},P_{\bar{b}_2+1,\bar{a}_1},P_{\bar{b}_1+1,\bar{a}_1}P_{\bar{b}_1+1,1}|2]\end{aligned}\\
    \bar{N}_2&=[\bar{b}_1 \, \bar{b}_1+1]\dotsm [\bar{b}_{k+1}\, \bar{b}_{k+1}+1]\\
    \bar{N}_3&=\spaa[\bar{a}_1+1\,\bar{a}_1]\dotsm \spaa[\bar{a}_k+1\,\bar{a}_k]\\
    \bar{D}_1&=P_{\bar{b}_1+1,1}^2P_{\bar{b}_1+1,\bar{a}_1}^2P_{\bar{b}_2+1,\bar{a}_1}^2\dotsc P_{\bar{b}_k+1,q+1}^2\\
    \bar{D}_2&=\bar{Z}_{2,q+1}Z_{q+2,1}\\
    \bar{D}_3&=\lan{1|P_{\bar{b}_1+1,1}|\bar{b}_1][\bar{b}_1+1|P_{\bar{b}_1+1,\bar{a}_1}|\bar{a}_1+1 \rangle\langle \bar{a}_1|P_{\bar{b}_2+1,\bar{a}_1}|\bar{b}_2]\dotsc [\bar{b}_k+1| P_{\bar{b}_k+1,q+1}|q+2}\ran ,
  \end{split}
\end{align}
with
\begin{equation}
  Z_{i,j}=\spaa[i\,i+1]\dotsm\spaa[j-1\,j],\qquad \bar{Z}_{i,j}=[i\,i+1]\dotsm[j-1\,j].
\end{equation}
\end{subequations}
}
The split-helicity form factor is then the sum of all recursion terms, or equivalently the sum of all possible MHV and \MHVbar zig-zags,  which is equal to
\begin{equation}
\label{eq:Fsolu}
F_{q-2,n-q-2} = \sum_{\{a_i,b_i\}}\frac{N_1N_2N_3}{D_1D_2D_3}+\sum_{\{\bar{a}_i,\bar{b}_i\}}\frac{\bar{N_1} \bar{N_2} \bar{N_3}}{\bar{D_1} \bar{D_2} \bar{D_3}}.
\end{equation}
Notice that for the form factors with equal number of negative and positive helicity gluons, the \MHVbar zig-zags can be obtained from the MHV ones by changing $(2,3,\dotsc,q) \to (1,n,\dotsc,q+1)$ and $\ang{ij}\to\sqr{ji}$.

Let us now explain the precise relation between the zig-zag diagrams and the paths on the split-helicity form factor lattice. Let a path with $r_1$ steps to the right, $l_1$ steps to the left followed by $r_2$ steps to the right etc. be represented by
\begin{equation}
    R^{r_k}\dotsm R^{r_2}L^{l_1}R^{r_1}.
\end{equation}
Then an MHV zig-zag labelled by $\{a_i,b_i\}$ corresponds to the path:
\begin{equation*}
L^{a_1-1}R^{b_1-b_2}\dotsm L^{a_{k}-a_{k-1}}R^{b_{k}-b_{k+1}}L^{q-1-a_{k}}R^{b_{k}-(q+1)},
\end{equation*}
while an \MHVbar zig-zag labelled by $\{\bar{a}_i,\bar{b}_i\}$ corresponds to the path:
\begin{equation*}
  R^{\bar{a}_1+1}L^{\bar{b}_2-\bar{b}_1}\dotsm R^{\bar{a}_{k}-\bar{a}_{k-1}}L^{\bar{b}_{k+1}-\bar{b}_{k}}R^{\bar{a}_{k}-q-1}L^{q-\bar{b}_{k+1}-1}\, .
\end{equation*}
Note that  if there are no $a_i$ indices in the
MHV zig-zag diagram we set $a_1=1$; and if there are no ${\bar a_i}$
in the $\overline{\rm MHV}$ zig-zag diagram we set ${\bar a}_1=n$.
All powers in the above formulae are modulo $n$.


\subsubsection{Examples}
Here we present some examples to show that the solution (\ref{eq:Fsolu}) reproduces the correct expressions.

\noindent
{\bf MHV case}

\noindent
The zig-zag diagrams collapse onto a point between $1$ and $2$ as there are neither $b_i$ nor $\bar{a}_i$. Hence, the only contributions are $N_1=\spaa[12]$ and $D_2=F_{2,1}$ and
\begin{equation}
F_{1,n-3}(1_\phi,2_\phi,3^+,\ldots,n^+) = \frac{\spaa[12]}{\spaa[23]\spaa[34] \ldots \spaa[n1]},
\end{equation}
as required. The situation for MHV amplitudes is similar \cite{BFRSV06}.
An equivalent calculation for the \MHVbar zig-zag gives the \MHVbar form factor.

\noindent
{\bf NMHV case}

\noindent
At four points, there is exactly one MHV and one \MHVbar zig-zag, representing one move to the left and one move to the right. Comparing with equations (\ref{eq:zzh}) and (\ref{eq:zzah}) one can read off $b_1=4$ for the MHV zig-zag and $\bar{b}_1=2$ for the \MHVbar zig-zag.
\begin{equation}
  \begin{minipage}{60pt}
  \begin{tikzpicture}
    \node (F) at (0,1) {$F_{1,1}$};
    \node (F2) at (-0.1,1) {\phantom{$F_{1,1}$}};
    \draw (0,-1) -- (1,0) (-1,0) -- (0,-1);
    \draw (-1,0) to (F);
    \draw (1,0) to (F);
    \draw[very thick,red](F2) to(-1.05,0.05);
  \end{tikzpicture}
  \end{minipage}
  =
  \begin{minipage}{72pt}
    \begin{tikzpicture}
    \draw                   (0,0) circle (1);
    \filldraw[fill=gray, draw=black]    (135:1) circle (3pt) node[left]{2}
                        (-135:1) circle (3pt) node[left]{1};
    \filldraw[fill=white, draw=black]   (45:1)  circle (3pt) node[right]{3};
    \filldraw[black]            (-45:1) circle (3pt) node[right]{4};
    \draw[red]              (180:1) .. controls (180:0.75) .. (-90:1) -- (90:1); 

\end{tikzpicture}
  \end{minipage}
  = \frac{[24]^2}{[32][43]}\frac{\langle 1| q | 4]}{\langle 1 |q|2]}\frac{1}{s_{234}}
\end{equation}
\begin{equation}
  \begin{minipage}{60pt}
  \begin{tikzpicture}
    \node (F) at (0,1) {$F_{1,1}$};
    \node (F2) at (0.1,1) {\phantom{$F_{1,1}$}};
    \draw (0,-1) -- (1,0) (-1,0) -- (0,-1);
    \draw (-1,0) to (F);
    \draw (1,0) to (F);
    \draw[very thick,blue](F2) to(1.05,0.05);
  \end{tikzpicture}
  \end{minipage}
  =
  \begin{minipage}{72pt}
    \begin{tikzpicture}
    \draw                   (0,0) circle (1);
    \filldraw[fill=gray, draw=black]    (135:1) circle (3pt) node[left]{2}
                        (-135:1) circle (3pt) node[left]{1};
    \filldraw[fill=white, draw=black]   (45:1)  circle (3pt) node[right]{3};
    \filldraw[black]            (-45:1) circle (3pt) node[right]{4};
    \draw[blue]             (180:1) .. controls (180:0.75) .. (90:1) -- (-90:1); 

\end{tikzpicture}
  \end{minipage}
  = \frac{\ang{13}^2}{\ang{34}\ang{41}}\frac{\langle 3| q | 2]}{\langle 1 |q|2]}\frac{1}{s_{341}}
\end{equation}
This result is in agreement with the previous section.

\begin{figure}[h]
\centering
\begin{equation*}
\begin{minipage}{120bp}
\begin{tikzpicture}
        \node (F) at (-2,4) {$F_{1,n-3}$};
        \node (F2) at (-1.9,4) {\phantom{F}};
        \node (F3) at (-2.1,4) {\phantom{F}};

    \draw (-1,1) -- (0,0) -- (1,1) -- (0,2)-- (-1,1) (0,2) -- (-1,3) -- (-2,2) -- (-1,1);
    \draw[very thick, red] (F) to (-3,3);

    \draw[very thick, blue] (-1.1,3) -- (-2.05,2.05);
    \draw[dashed, blue] (F3) to (-1.1,3);

    \draw[very thick,green] (-0.9,3) -- (0.1, 2)-- (-0.95,0.95);
    \draw[dashed, green] (F2) to (-0.9,3);

    \draw[dashed] (-2,2) -- (-3,3) (-1,3) ;
        \draw[dashed] (-1,3) to (F);

\end{tikzpicture}
\end{minipage}
=
\begin{minipage}{150bp}
\begin{tikzpicture}
    \draw (1.5,-1) -- (0,-1) arc (270:90:1) -- +(3.5,0)arc (90:-90:1) -- +(-1.5,0);
        \draw[dotted] (1.5,-1) -- (2,-1);
        \filldraw[fill=gray, draw=black]  (150:1) circle (3pt) node[left]{2};
        \filldraw[fill=gray, draw=black]  (210:1) circle (3pt) node[left]{1};
    \filldraw[black] 
                            (0,-1) circle (3pt) node[below=4pt]{$n$}
                (1,-1) circle (3pt) node[below=4pt]{$n-1$}
                (2.5,-1) circle (3pt) node[below=4pt]{$5$}
                (3.5,-1) circle (3pt) node[below=4pt]{$4$};
    \filldraw[fill=white, draw=black]
                (3.5,1) circle (3pt) node[above=4pt]{$3$};
    \draw[green] (-1,0) .. controls (0,0) .. (225:1) -- (1.75, 1);
    \draw[blue] (-1,0) .. controls (0,0) .. (0.5,-1) -- (1.75, 1);
    \draw[red] (-1,0) .. controls (2,0) .. (3,-1) -- (1.75, 1);
\end{tikzpicture}
\end{minipage}
\end{equation*}
\caption{\it Correspondence of lattice paths and MHV zig-zags for NMHV form factors.}
\label{fig:NMHVfig}
\end{figure}
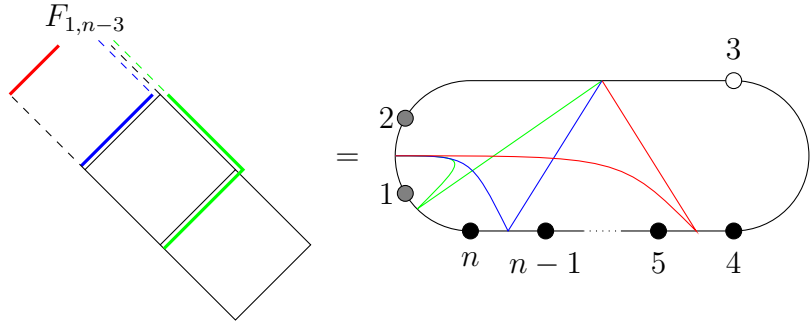

In general, for the NMHV form factors, there is one \MHVbar zig-zag corresponding to the path which proceeds along the $\rm{NMHV}$ line until it reaches the $\overline{\rm{MHV}}$ edge of the lattice, and $n-3$ MHV zig-zags where the path shifts onto the $\rm{MHV}$ edge before it arrives at the \MHVbar edge.
The MHV paths and the corresponding zig-zags are shown in Figure \ref{fig:NMHVfig}.

\noindent
{\bf An N$^2$MHV example}

\noindent
As it can be seen from the lattice in Figure \ref{lattice}, there are three MHV and three \MHVbar terms in the recursion of the six-point split-helicity form factor. These are listed below, where the subscripts encode the shape of the path as described earlier. For example, $F_{RLL}$ is the term which corresponds to the path that starts with a step to right and terminates at the MHV edge with two steps to the left.
The MHV terms are:
\begin{itemize}
\begin{subequations}
\item $b_1=5$, no $a$:
\begin{equation}
  F_{LL}= \begin{minipage}{75pt}\begin{tikzpicture}
  \draw (0,0) circle (1);
  \filldraw[fill=gray, draw=black]  (150:1) circle (3pt) node[left]{2};
  \filldraw[fill=gray, draw=black]  (210:1) circle (3pt) node[left]{1};
  \filldraw[fill=white, draw=black] (90:1)  circle (3pt) node[above]{3};
  \filldraw[fill=white, draw=black] (30:1)  circle (3pt) node[right]{4};
  \filldraw[fill=black, draw=black] (-30:1) circle (3pt) node[right]{5};
  \filldraw[fill=black, draw=black] (-90:1) circle (3pt) node[below]{6};
  \draw (180:1).. controls (180:0.75) ..  (-60:1) -- (60:1);
\end{tikzpicture}\end{minipage} = - \frac{\sqr{25}^2}{\sqr{23}\sqr{34}\sqr{45}\ang{61}}\frac{1}{P_{2,5}^2}\frac{[5|P_{2,4}|1\rangle}{[2|P_{2,5}|6\rangle}
\end{equation}

\item $b_1=6$, no $a$:
\begin{equation}
  F_{RLL}=\begin{minipage}{75pt}\begin{tikzpicture}
  \draw (0,0) circle (1);
  \filldraw[fill=gray, draw=black]  (150:1) circle (3pt) node[left]{2};
  \filldraw[fill=gray, draw=black]  (210:1) circle (3pt) node[left]{1};
  \filldraw[fill=white, draw=black] (90:1)  circle (3pt) node[above]{3};
  \filldraw[fill=white, draw=black] (30:1)  circle (3pt) node[right]{4};
  \filldraw[fill=black, draw=black] (-30:1) circle (3pt) node[right]{5};
  \filldraw[fill=black, draw=black] (-90:1) circle (3pt) node[below]{6};
  \draw (180:1) .. controls (180:0.75) ..  (240:1) -- (60:1);
\end{tikzpicture}\end{minipage} = \frac{1}{\ang{45}\ang{56}\sqr{23}}\frac{1}{P_{2,6}^2P_{4,6}^2}\frac{\ang{1|P_{2,6}P_{4,6}|4}[2|P_{4,6}|4\rangle^2}{[2|P_{2,6}|1\rangle \langle 6 |P_{4,6}|3]}
\end{equation}

\item $b_1=6$, $b_2=5$, $a_1=2$:

\begin{equation}
  F_{LRL}= \begin{minipage}{75pt}\begin{tikzpicture}
  \draw (0,0) circle (1);
  \filldraw[fill=gray, draw=black]  (150:1) circle (3pt) node[left]{2};
  \filldraw[fill=gray, draw=black]  (210:1) circle (3pt) node[left]{1};
  \filldraw[fill=white, draw=black] (90:1)  circle (3pt) node[above]{3};
  \filldraw[fill=white, draw=black] (30:1)  circle (3pt) node[right]{4};
  \filldraw[fill=black, draw=black] (-30:1) circle (3pt) node[right]{5};
  \filldraw[fill=black, draw=black] (-90:1) circle (3pt) node[below]{6};
  \draw (180:1) .. controls (180:0.75) .. (240:1) -- (120:1) -- (-60:1) -- (60:1);
\end{tikzpicture}\end{minipage} =\frac{1}{\sqr{34}\sqr{45}}\frac{1}{P_{2,6}^2P_{3,6}^2P_{3,5}^2} \frac{\langle 1 |P_{2,6}P_{3,6}P_{3,5}|5]\sqr{2|P_{3,6}P_{3,5}|5}^2}{[2|P_{2,6}|1\rangle\langle 6 |P_{3,6}1|2][3|P_{3,5}|6\rangle }
\end{equation}
\end{subequations}
\end{itemize}
The \MHVbar terms are:
\begin{itemize}
\begin{subequations}
\item $\bar{b}_1=3$, no $\bar{a}$
\begin{equation}
F_\mathrm{RR}=\begin{minipage}{75pt}\begin{tikzpicture}
  \draw (0,0) circle (1);
  \filldraw[fill=gray, draw=black]  (150:1) circle (3pt) node[left]{2};
  \filldraw[fill=gray, draw=black]  (210:1) circle (3pt) node[left]{1};
  \filldraw[fill=white, draw=black] (90:1)  circle (3pt) node[above]{3};
  \filldraw[fill=white, draw=black] (30:1)  circle (3pt) node[right]{4};
  \filldraw[fill=black, draw=black] (-30:1) circle (3pt) node[right]{5};
  \filldraw[fill=black, draw=black] (-90:1) circle (3pt) node[below]{6};
  \draw (180:1) .. controls (180:0.75) .. (60:1) -- (-60:1);
\end{tikzpicture}\end{minipage} =\frac{\ang{14}^2}{\ang{45}\ang{56}\ang{61}\sqr{23}}\frac{1}{P_{4,1}^2}\frac{\langle 4|P_{4,1}|2]}{\langle 1 |P_{4,1}|3]}
\end{equation}
\item $\bar{b}_1=2$, no $\bar{a}$
\begin{equation}
 F_\mathrm{LRR}=\begin{minipage}{75pt}\begin{tikzpicture}
  \draw (0,0) circle (1);
  \filldraw[fill=gray, draw=black]  (150:1) circle (3pt) node[left]{2};
  \filldraw[fill=gray, draw=black]  (210:1) circle (3pt) node[left]{1};
  \filldraw[fill=white, draw=black] (90:1)  circle (3pt) node[above]{3};
  \filldraw[fill=white, draw=black] (30:1)  circle (3pt) node[right]{4};
  \filldraw[fill=black, draw=black] (-30:1) circle (3pt) node[right]{5};
  \filldraw[fill=black, draw=black] (-90:1) circle (3pt) node[below]{6};
  \draw (180:1) .. controls (180:0.75) ..  (120:1) -- (-60:1);
\end{tikzpicture}\end{minipage} =\frac{1}{\sqr{34}\sqr{45}\ang{61}}\frac{1}{P_{3,5}^2P_{3,1}^2}\frac{\langle 1|P_{3,5}|5]^2\sqr{5|P_{3,5}P_{3,1}|2}}{\langle 1| P_{3,1} |2][3|P_{3,5}|6\rangle }
\end{equation}
\item $\bar{b}_1=2$, $b_2=3$, $\bar{a}_1=6$
\begin{equation}
  F_\mathrm{RLR}=\begin{minipage}{75pt}\begin{tikzpicture}
  \draw (0,0) circle (1);
  \filldraw[fill=gray, draw=black]  (150:1) circle (3pt) node[left]{2};
  \filldraw[fill=gray, draw=black]  (210:1) circle (3pt) node[left]{1};
  \filldraw[fill=white, draw=black] (90:1)  circle (3pt) node[above]{3};
  \filldraw[fill=white, draw=black] (30:1)  circle (3pt) node[right]{4};
  \filldraw[fill=black, draw=black] (-30:1) circle (3pt) node[right]{5};
  \filldraw[fill=black, draw=black] (-90:1) circle (3pt) node[below]{6};
  \draw (180:1) .. controls (180:0.75) ..  (120:1) -- (240:1) -- (60:1) -- (-60:1);
\end{tikzpicture}\end{minipage} =\frac{1}{\ang{45}\ang{56}}\frac{1}{P_{3,1}^2P_{3,6}^2P_{4,6}^2}\frac{\langle 4 | P_{4,6} P_{3,1}|1\rangle^2\langle 5 | P_{4,6}P_{3,6}P_{3,1}|2]}{\langle 1 |P_{3,1}|2][3|P_{3,6}|1\rangle\langle 6 |P_{4,6}|3][4|P_{4,6}|6\rangle}
\end{equation}
\end{subequations}
\end{itemize}
We have checked this result against an MHV diagram calculation and both methods yield the same result.


\section{Supersymmetric form factors and Ward identities \label{section-super}}

The purpose of this section is to write down supersymmetric Ward identities for certain appropriately defined form factors of supersymmetric operators. By solving these Ward identities, we will learn about the structure of these form factors.

To begin, we recall the familiar fact that  in $\cN=4$ SYM one can efficiently package all scattering amplitudes
with fixed total helicity and fixed number of particles $n$ into a superamplitude \cite{Nair}, thereby making manifest some of the supersymmetries of the theory. This object depends on auxiliary fermionic variables $\eta_{i, A}$, one for each particle  $i=1, \ldots , n$,   with $A$ an anti-fundamental $SU(4)$ index. The superamplitude can be Taylor-expanded in the $\eta$ variables, with a specific correspondence between powers of $\eta$ and particular external states. This correspondence  can be read off from the Nair super-wavefunction \cite{Nair}, which encodes all the annihilation operators of the physical states,
\beq
\label{superNair}
\Phi(p,\eta) := g^+(p) + \eta_A \lambda^A(p) + {\eta_A \eta_B\over
2!} \phi^{AB}(p) + \epsilon^{ABCD} {\eta_A \eta_B \eta_C\over 3!}
\bar\lambda_D(p) + \eta_1 \eta_2 \eta_3 \eta_4 g^-(p)
\, ,
\eeq
where $(g^+(p), \ldots , g^{-}(p))$ are the annihilation operators of the corresponding states.
In order to select a state with a particular
helicity $h_i$, we need to expand the superamplitude and pick the term with $2 - 2h_i$ powers of  $\eta_i$.

This familiar framework becomes richer for form factors. Indeed, one can consider form factors of bosonic operators -- such as
$\Tr ( \phi_{AB} \phi_{AB})$ -- with an external supersymmetric state described using the Nair approach, but one can also supersymmetrise the operator itself, as we shall see in the next section.

A comment on notation -- we denote a  form factor  as
$\lan 0 | \Phi(1) \cdots \Phi (n)\, \cO \ | 0  \ran$ or equivalently $\lan 1\cdots n | \, \cO \ | 0  \ran$,  where $|i \ran := \Phi^\dagger (i) |0\ran $ is a Nair superstate, which satisfies
\beq
\label{eigen}
\lan \, i\,  |\,  P  = \lan \, i\,  |\,   p_i  \ , \qquad \lan \, i\,  |\,  Q   = \lan \, i\,  |\,   \lambda_i \eta_i  \ ,
\qquad
\lan \, i\,  |\,  \bar{Q}  = \lan \, i\,  |\,   \,  {{\partial} \over \partial \eta_i} \, \tilde{\lambda}_i \ ,
\eeq
where  the derivative in the last equation acts on the state on its left.
We also adopt the notation $\lan1 \cdots n | := \lan 0 | \Phi(1) \cdots \Phi(n)$.

\subsection{Form factor of the chiral stress-tensor multiplet operator}

We now consider the form factor of the chiral supersymmetric operator%
\footnote{
A quick reminder of harmonic superspace \cite{harm1,harm2}  conventions, following closely
\cite{Eden:2011yp,Eden:2011ku}.
We introduce the harmonic projections of  the $\theta^A_{\alpha}$ and $\bar{\theta}_A^{\dot\alpha}$ superspace coordinates and
of the supersymmetry charges $Q^{\alpha}_A$,  $\bar{Q}_{\dot\alpha,}^A$
as $\theta_{\alpha}^{+ a} := \theta^A_{\alpha} u_A^{+ a}$,   $\bar{\theta}^{\dot\alpha}_{- a} :=\bar{ \theta}^{\dot\alpha}_{A} \bar{u}^A_{- a}$, and
$Q_{\pm a}^{\alpha} := \bar{u}^A_{\pm a} Q^\alpha_A$, $\bar{Q}^{+a}_{\dot\alpha} := u_A^{+ a} \bar{Q}_{\dot \alpha}^{A}$
with the
harmonic $SU(4)$  $u$ and $\bar{u}$ normalised as in Section 3 of
\cite{Eden:2011yp}.
}
$\cT (x, \theta^{+})$ considered recently  in
\cite{Eden:2011yp, Eden:2011ku}. This  operator  is the chiral part of the stress-tensor multiplet operator,
$\cT (x, \theta^{+}) :=  \cT (x, \theta^{+}, \bar\theta_{-} = 0,  u )$ and we report here its expression from  \cite{Eden:2011yp}  for convenience,
\beqa
\label{texp}
\cT (x, \theta^+) &=& \Tr ( \phi^{++} \phi^{++}) + i 2 \sqrt{2} \theta_{\alpha}^{+a} \Tr ( \lambda_a ^{+ \alpha}  \phi^{++})
\nonumber \\
&+& \theta_{\alpha}^{+a} \eps_{ab} \theta^{+b}_{\beta} \Tr \Big( \lambda^{+c ( \alpha} \lambda_{c}^{ + \beta )} - i \sqrt{2} F^{\alpha \beta} \phi^{++} \Big)
\nonumber \\
&-& \theta_{\alpha}^{+a} \eps^{\alpha \beta} \theta_{\beta}^b \Tr \Big( \lambda^{+ \gamma}_{(a} \lambda^{+}_{b) \gamma} - g \sqrt{2} [ \phi^{+C}_{(a} , \bar{\phi}_{C \, +b)} ] \phi^{++} \Big)
\nonumber \\
&-& {4\over3} ( \theta^{+}  )^{3\, a}_{\alpha} \Tr \Big( F_{\beta}^{\alpha} \lambda_{a}^{+ \beta} + i g [ \phi_a^{+B} ,  \bar{\phi}_{BC} ] \lambda^{C \alpha}\Big) + {1\over 3} (\theta^+)^4 \,  \cL
\ .
\eeqa
Notice that the $(\theta^+)^0$ component is nothing but the scalar operator $\Tr (\phi^{++} \phi^{++})$, whereas the $(\theta^+)^4$ component is the on-shell Lagrangian.

Next we describe how to use supersymmetric Ward identities in order to  constrain  form factors, slightly extending the usual procedure for amplitudes.  Ward identities associated with a certain symmetry generator $s$ which leaves the vacuum invariant are obtained in a standard way  \cite{Grisaru:1976vm,Grisaru:1977px,Mangano:1990by, Elvang:2010xn}
by expanding the identity
\beq
0 \ = \ \lan 0 | [s \, ,   \Phi(1) \cdots \Phi (n)\, \cO \,] | 0  \ran
\ ,
\eeq
or
\beq
0 \ = \   \lan 0 |  \Phi(1) \cdots \Phi (n)\,  [s \, , \, \cO] \,| 0  \ran \, + \, \sum_{i=1}^n  \lan 0 | \,  \Phi(1) \cdots [s \, , \, \Phi(i) ] \cdots \Phi (n)\, \cO | 0  \ran
\ .
\eeq
For instance, by considering $s$ to be the  momentum generator $\cP$ and using
$[\cP_\mu, \cO (x) ] = -i \partial_{\mu} \cO(x)$ as well as the first equation of \eqref{eigen},
we obtain
\beq
-i\, \lan 0 |  \Phi(1) \cdots \Phi (n)\,  \partial_{\mu}  \cO (x)  \,  | 0  \ran + ( \sum_{i=1}^n p_i )\lan 0 |     \Phi(1) \cdots \Phi (n)\, \cO (x) | 0 \ran = 0 \ .
\eeq
Fourier transforming $x$ to $q$ and integrating by parts one obtains
\beq
(q - \sum_{i=1}^n p_i) F(q; 1, \ldots , n) = 0
\ ,
\eeq
where
\beq
\label{fnonsusydef}
F(q; 1, \ldots , n) \ := \
\int\!\!d^4x \ e^{-i q x} \ \lan  1 \cdots n |   \cO (x)  \,  |  0 \ran\ .
\eeq
From this it follows that
\beq
\label{sopra}
F(q; 1, \ldots , n) \ = \ C \cdot \delta^{(4)} (q - \sum_{i=1}^n p_i)
\ .
\eeq
$ C$  can be fixed by further integrating
both sides of \eqref{sopra} with a $d^4q$ measure and using  \eqref{fnonsusydef}, which leads to
$C  =  \lan 0 | \Phi(1) \cdots \Phi (n)\,   \cO (0)  \,  | 0 \ran\  =\lan 1 \cdots n  |   \cO (0)  \,  |0\ran\ $.

Similarly, we now consider  Ward identities for the  harmonic projections $Q_{\pm a}^{\alpha}$, $a=1,2$, of the $Q$-supersymmetry generators.
We obtain
\beq
\label{wipm} 0
\ = \
\lan 0 |\, \Phi(1)
\cdots \Phi (n) [Q_{\pm}  \, , \, \cT (x, \theta^{+} ) ] \,  | 0
\ran \, + \, \sum_{i=1}^n  \lan 0 |  \,  \Phi(1) \cdots [Q_{\pm} \,
, \, \Phi(i) ] \cdots \Phi (n)\, \cT (x, \theta^{+} )\, | 0  \ran \
.
\eeq
We now have to discuss how supersymmetry acts on the chiral part of  $\cT(x, \theta^+)$ as well as on the states.

In general the supersymmetry algebra closes only up to gauge transformations and equations of motion,%
\footnote{We would like to thank Paul Heslop for a useful conversation on these issues.}
however we consider here gauge-invariant operators such as $\cT$ which, furthermore, are made only of a subset of all fields, namely $\phi^{AB}$, $\lambda^A_\alpha$ and $F_{\alpha \beta}$.
It is an  important  fact that the algebra of the $Q$-generators closes off shell  on the chiral part of $\cT$ \cite{Eden:2011yp}, and hence these generators can be realised as differential operators. Of course,  representing the $\bar{Q}$-generators in terms of differential operators is, in general, problematic, because the full supersymmetry algebra closes only on shell.

Moreover,  for the chiral operator $\cT(x, \theta^{+})$ we have broken $\bar{Q}^{-}$ since we have set $\theta^{-} =0$ and hence we do not have a representation for this operator. For the $Q_{\pm}$-variation of $\cT(x, \theta^{+})$ we have,
\beq
\label{two} [Q_{-}  \, , \, \cT (x, \theta^{+} ) ] \ = \ 0
 \ , \qquad
 [Q_{+}  \, , \, \cT (x, \theta^{+} ) ] \ = \  i {\partial \over \partial \theta^{+} } \cT (x, \theta^{+} )  \ .
 \eeq
Note that since we consider the chiral part of the stress-tensor multiplet we have set $\bar{\theta}=0$ and hence we have dropped $\bar{\theta}$ dependent terms in the realisation
of $Q$ and $\bar{Q}$.
Then the first relation is obvious since $ \cT (x, \theta^{+} )$ is
independent of $\theta^{-}$. This also makes manifest the fact that all component operators
of $\cT(x, \theta^{+})$ are annihilated by $Q_{-a}^{\alpha}$ \cite{Eden:2011yp}. On
the other hand,  $Q_{+a}^{\alpha}$ relates
different components of the supermultiplet, as the second relation
in \eqref{two} shows.

We define the super form factor as the  super Fourier transform  of
the matrix element $\lan 1 \cdots n   |  \cT (x, \theta^{+} )  \,  |
0 \ran$, i.e.
\beq
\label{sft22}
\cF_{\cT} (q, \gamma_{+}; 1, \ldots, n)
\ := \ \int\!\!d^4x\, d^4 \theta^{+} \
e^{-(iqx + i \theta_\alpha^{+a}
\gamma_{+a}^\alpha) } \, \langle  \, 1 \cdots n\,  | \cT(x, \theta^+)
\,  | 0   \ran\ \, ,
\eeq
where $\gamma_{+a}^\alpha $ is the  Fourier-conjugate  variable to $\theta^{+a}_\alpha $. Note that there
is no $\gamma_{-a}^\alpha $  variable, since $\theta^{-a}_\alpha $
has been set to zero in order to define the chiral part of the
stress-tensor multiplet. The Ward identities \eqref{wipm} can then
be recast as
\beqa \label{fin}
 \big( \sum_{i=1}^n \lambda_i \eta_{-,
i} \big) \ \cF_{\cT} (q, \gamma_{+}; 1, \ldots, n) & = & 0\ ,
\nonumber \\
\big( \sum_{i=1}^n \lambda_i \eta_{+, i} \,  -  \, \gamma_{+} \big)
\ \cF_{\cT} (q, \gamma_{+}; 1, \ldots, n) & = & 0 \, ,
\eeqa
where we have also introduced
\beq
\eta_{\pm a, i} \, := \, \bar{u}_{\pm a}^A \eta_{A, i}\ .
\eeq
In
arriving at \eqref{fin} we have used \eqref{two} as well as the
second relation in \eqref{eigen}. Next, we observe that \eqref{fin}
are solved by \beq \label{sff1bis}
\cF_{\cT} (q, \gamma_{+}; 1, \ldots, n)
\ = \ \delta^{(4)} (q - \sum_{i=1}^n \l_i \lt_i)  \, \delta^{(4)}
\big( \gamma_{+} -  \sum_{i=1}^n \eta_{+, i} \l_i   \big)\,
\delta^{(4)} \big( \sum_{i=1}^n \eta_{-, i}  \l_{i}  \big) \, R,
\eeq
for some function $R$ which in principle depends on all  bosonic and fermionic variables.
The simplest example is that of  the MHV form factor, where
the function  $R$ has a particularly simple expression derived in \cite{bsty2}, namely
\beq
\label{rfmhv}
R^{\rm MHV} \ = \ {1\over \lan12\ran \cdots \lan n1 \ran} \ .
\eeq
Notice that for an $\mathrm{N}^k\mathrm{MHV}$ form factor, $R$ has fermionic degree $4k$.

We can further constrain  $R$ by using some of the $\bar{Q}$-supersymmetries. More precisely, an inspection of the supersymmetry transformations of the fields reveals that a  $\bar{Q}^{-} $ transformation on the chiral part of the stress-tensor multiplet produces operators which are part of the full stress-tensor multiplet but not of its chiral truncation.  Also,  since $[Q_-, \cT(x, \theta^+)] = 0 $ we cannot realise $\bar{Q}^{-}$ such that its anticommutator with $Q_-$ gives a translation.  One could of course still write a  Ward identity for $\bar{Q}^{-}$, but this would involve operators of the full multiplet.

On the other hand, the $\bar{Q}^+$-supersymmetry charge moves in the opposite direction of $Q_{+}$ across the different components of $\cT(x, \theta^+)$,  and is therefore realised as
$\bar{Q}_{\dot\alpha}^+ = - \theta^{+\,  \alpha} {\partial /
\partial x^{\dot \alpha \alpha}}$.

We should stress at this point that the supersymmetry algebra on
component fields closes only up to equations of motion and gauge
transformations (the latter drop out since we consider gauge
invariant operators). An important exception is the subalgebra
formed by the $Q$'s alone which does close off-shell for the fields
appearing in $\cT(x, \theta^+)$ \cite{Eden:2011yp}. Now we use the
fact that matrix elements of terms proportional to equations of
motion vanish at tree level, to argue that for our tree-level form
factors the algebra formed by $Q_+$ and $\bar{Q}^+$ does close and,
therefore, can be realised in the fashion described above. Thus, we
can consider the $\bar{Q}^{+} $ Ward identity, which gives, after
integrating by parts and using the third relation of \eqref{eigen},
\beq \label{chiralQbar} \big( \sum_{i=1}^n \tilde{\lambda}_i {\partial
\over \partial \eta_{+, i}} \, - \, q {\partial \over \partial
\gamma_{+}} \big) \cF_{\cT} (q, \gamma_{+}; 1, \ldots, n) \ = \ 0 \
. \eeq
Acting on \eqref{sff1bis}, we obtain the following relation
for $R$,
\beq \label{rcon} \delta^{(4)} (q - \sum_{i=1}^n \l_i
\lt_i) \, \delta^{(4)}\big( \gamma_{+} -  \sum_{i=1}^n \eta_{+, i}
\l_i \big)\, \delta^{(4)} \big( \sum_{i=1}^n \eta_{-, i}  \l_{i}
\big)\ \left[ \big( \sum_{i=1}^n \tilde{\lambda}_i {\partial \over
\partial \eta_{+, i}} \, - \, q {\partial \over \partial \gamma_{+}}
\big) \, R \right]
 \ = \ 0
\ .
\eeq
Notice that \eqref{rcon}  implies a realisation of the  supersymmetry generators on the form factor as
\beq
Q^{\alpha}_{ +a} \ = \  \sum_{i=1}^n \lambda_{i}^{ \alpha} \eta_{+ a, i}  \ - \   \gamma_{+a}^{ \alpha}\ , \qquad
Q^{\alpha}_{ -a} \ = \  \sum_{i=1}^n \lambda_{i}^{ \alpha} \eta_{- a, i}  \ ,
\eeq
whereas for $\bar{Q}_{\dot\alpha}^{+a}$,
\beq
\bar{Q}_{\dot\alpha}^{+a} \ = \ \sum_{i=1}^n \tilde\lambda_{i, \dot\alpha} {\partial \over\partial  \eta_{+ a, i}}  \ - \  q_{\alpha\dot\alpha }{\partial \over \partial \gamma_{\alpha\, +a}}
\ .
\eeq

\subsection{Examples}
In the previous section we have derived the general form of the supersymmetric form factor  defined in \eqref{sft22}. This expression
is given in \eqref{sff1bis}, and was obtained  by solving Ward identities related to  translations and $Q_{\pm}$-supersymmetries. The use of $\bar{Q}^{+} $ supersymmetry   led to the constraint  \eqref{rcon} on the function $R$.
For the sake of illustration, we now present a few examples of component form factors derived from \eqref{sff1bis}.

\subsubsection{Form factor of $\Tr
(\phi^{++} \phi^{++})$}

\noindent
Our first example is the form factor of
$\Tr (\phi^{++} \phi^{++})$,  which appears as the $(\theta^+)^0$-term in the expansion of
$\cT(x, \theta^+) $ in \eqref{texp}. In this case, since
\beq \int\!d^4 \theta^{+} \ e^{i
\theta_\alpha^{+a} \gamma_{+a}^\alpha }  \ = \  (\gamma_{+})^4\ ,
\eeq
we need to extract the $(\gamma_{+})^4$ component of
\eqref{sff1bis}. This gives
\beq \int\!d^4x  \ e^{-iqx } \, \langle 1
\cdots   n | \Tr (\phi^{++} \phi^{++})(x)  |0   \rangle\  = \
 \delta^{(4)} (q - \sum_{i=1}^n \l_i \lt_i)  \, \delta^{(4)} \big( \sum_{i=1}^n \eta_{-a, i}  \l_{i}^{\alpha}  \big)
\, R \ , \eeq or \beq \label{sopff} \langle 1 \cdots   n | \Tr
(\phi^{++} \phi^{++})(0)  | 0  \rangle\  = \
 \delta^{(4)} \big( \sum_{i=1}^n \eta_{-a, i}  \l_{i}^{\alpha}  \big)
\, R
\ .
\eeq
Notice that with the help of  \eqref{sopff} we can rewrite the
supersymmetric form factor $\cF_{\cT} (q, \gamma_{+}; 1, \ldots, n)
$ as
\beq \label{sft} \cF_{\cT} (q, \gamma_{+}; 1, \ldots, n) \ = \
\delta^{(4)} (q - \sum_{i=1}^n \l_i \lt_i) \,  \delta^{(4)} \big(
\gamma_{+} -  \sum_{i=1}^n \eta_{+,i} \l_i   \big) \,  \langle 1
\cdots  n | \cT (0, 0) | 0 \rangle\ \, , \eeq
since $\cT(0 ,0) :=
\Tr  (\phi^{++} \phi^{++})(0) $.
In other words, the function $R$ appearing in the $\cT(x, \theta^{+})$ form factor can be calculated from the form factor of its lowest component%
\footnote{One could arrive at \eqref{sft}  in a much more
straightforward way by noticing that $\cT (x, \theta^{+a}_{\alpha})
= \exp (i \mathcal{P} x )   \exp (i Q_{+a}^{\alpha}
\theta^{+a}_{\alpha}  ) \cT(0, 0) \exp (- i  \mathcal{P} x ) \exp (-
i Q_{+a}^{\alpha} \theta^{+a}_{\alpha}  )$ and using the invariance
of the vacuum under supersymmetry and translations. } $
\Tr (\phi^{++}
\phi^{++})(0) $.
Similar considerations apply to form factors of other half BPS operators such as  $\Tr (\phi^{++})^n$ with $n>2$.

\subsubsection{Form factor of the on-shell Lagrangian}

As a second important example,  we now consider the form factor for the on-shell Lagrangian, whose expression is
\cite{Eden:2011yp}
\beqa
\label{lagon}
\mathcal{L}\ = \
\Tr
\Big[
-{1\over 2}
F_{\alpha \beta} F^{\alpha \beta} + \sqrt{2} g \lambda^{\alpha A} [ \phi_{AB}, \lambda^{B}_{\alpha}]
- {1\over 8} g^2 [\phi^{AB} , \phi^{CD}] [ \phi_{AB}, \phi_{CD}] \Big]
\ .
\eeqa
Notice that it contains the self-dual part of  $\Tr (F^2)$. The on-shell Lagrangian
appears as the $(\theta^{+})^4$ coefficient of the
expansion of $\cT(x, \theta^+)$ in \eqref{texp}. The corresponding  Fourier
transform gives
\beq \int\!d^4 \theta^{+} \ e^{-i \theta_\alpha^{+a}
\gamma_{+a}^\alpha } (\theta^{+})^4 \ = \ 1\ ,
\eeq
i.e.~we have to
take the $\cO (\gamma^0) $ component of \eqref{sff1bis}.
This is
simply
\beq
\label{idd}
\lan 1 \cdots n  | \cL (0)  | 0  \ran \, = \, \delta^{(8)} \big( \sum_{i=1}^n
\eta_i \l_i\big) \, \cdot R \, .
\eeq
It is interesting to note that for an MHV form factor, \eqref{idd} is formally identical to the tree-level MHV superamplitude, except for a delta function of momentum conservation which now imposes $\sum_i p_i =q$ rather than the usual momentum conservation of the particles. This allows us to make an interesting observation for the limit $q \to 0$ in which this form factor reduces simply to the correspond scattering amplitude. Actually, it turns out that any form factor with the on-shell Lagrangian $\cL$ inserted reduces to the
corresponding scattering amplitude in the $q \to 0$ limit, since the insertion of the action corresponds to differentiating the path-integral for the amplitude with respect to the coupling \cite{Intriligator:1998ig,Eden:1999kw,Eden:2000mv}.

Another observation is that for the case of a gluonic state with MHV helicity configuration, \eqref{idd} agrees with the Higgs plus  multi-gluon or ``$\phi$-MHV" amplitude considered in \cite{dgk}. Indeed, if we have a gluonic state, we can effectively replace the on-shell Lagrangian \eqref{lagon} with its first term, the square of the self-dual field strength.

\subsubsection{Why is the maximally non-MHV form factor so simple?}

The simplest tree-level form factor is the MHV form factor, e.g.
\beq
\label{mhvffsd}
\lan 1^+ 2^+    \cdots i^{-} \cdots j^{-} \cdots  (n-1)^+ n^+ | \, \Tr (F_{\rm SD}^2) (0)\, | 0 \ran \, = \,
 { \lan i j \ran^4   \over \lan 12 \ran  \lan 23 \ran  \cdots
\lan n~1 \ran}  \, . \eeq
Interestingly, there are non-MHV form factors whose expression is  also remarkably simple.
Consider for example that  of
the self-dual field strength with an all  negative-helicity gluons  state -- we refer to this  as
the ``maximally non-MHV" form factor. The result for this quantity  is \cite{dgk}
\bea
\label{mnmhv}
\lan 1^-  \cdots  n^-  | \, \Tr (F_{\rm SD}^2) (0) \, | 0 \ran \, = \,
 { q^4 \over [ 1~2 ][ 2~3 ]\cdots
[ n~1 ]} ~. \eea
In the following we wish to show that the simplicity of   \eqref{mnmhv}  is  determined by the supersymmetric
Ward identity discussed earlier, and is linked to that of the MHV super form factor \eqref{rfmhv}.

Recall from \eqref{sft}  that  the super form factor of the chiral part of the stress-tensor multiplet  $\cT( x, \theta^+)$ has the form
\bea
\label{ecf}
{\cal F}_\cT =  \delta^{(4)} (q - \sum_{i=1}^n \l_i \lt_i)\,
\delta^{(4)}\big( \gamma_{+} -  \sum_{i=1}^n
\eta_{+, i} \l_i   \big) \ {\cal F}_{\phi^2} ~,
\eea
where
\bea
 {\cal F}_{\phi^2} := \lan 1 \cdots n   | \, \Tr (\phi^{++}
\phi^{++})(0) \, | 0 \ran \, = \, \, \delta^{(4)} \big( \sum_{i=1}^n
\eta_{-, i}  \l_{i}  \big) \, R \ .
\eea
For the MHV helicity configuration, the function
$R^{\rm MHV}$ is given in  \eqref{rfmhv},
\bea
{\cal F}_{\phi^2}^{\rm MHV} \, = \,  {  \delta^{(4)} \big( \sum_{i=1}^n
\eta_{-, i}  \l_{i}  \big) \over \lan 12\ran \cdots \lan n1\ran}
\ .
 \eea
We can now use this fact  and perform a Grassmann Fourier transform in order to derive
the maximally non-MHV super form factor,
\bea
{\cal F}^{\textrm{N$^{\rm max}$MHV}}_{\phi^2} \, = \,
 \prod_{i=1}^n \int\!\!d^4 \tilde\eta_i\
e^{i\eta_{i,A} \tilde\eta_{i}^{A}}\
{ \delta^{(4)} \big( \sum_{i=1}^n
\tilde\eta_{i}^+ \tilde\l_{i} \big) \over [ 12 ] \cdots [ n1 ]} \ .
\eea
Thus, the maximally non-MHV super form factor for the chiral part of the stress-tensor multiplet is
\bea
{\cal F}^{\textrm{N$^{\rm max}$MHV}}_{\cT}  \, = \, \delta^{(4)} (q - \sum_{i=1}^n \l_i \lt_i)\, \, \delta^{(4)}\big(
\gamma_{+} -  \sum_{i=1}^n \eta_{+, i} \l_i   \big) \, {\cal
F}^{\textrm{N$^{\rm max}$MHV}}_{\phi^2} ~.
\eea
We now focus on the component corresponding to the self-dual field strength, which can be obtained
from the coefficient of $(\gamma_+)^0$. This is given by%
\footnote{In the following equation we omit a trivial delta function of momentum conservation.}
\bea
&& \delta^{(4)}\big(
\sum_{i=1}^n \eta_{+, i} \l_i   \big) \, \prod_{i=1}^n \int\!\!d^4
\tilde\eta_i \ e^{i\eta_i \tilde\eta_i} \ { \delta^{(4)} \big(
\sum_{i=1}^n \tilde\eta_{i}^+ \tilde\l_{i} \big) \over [ 12 ] \cdots
[ n1 ]}
\nonumber\\
&=&  \delta^{(4)}\big(
\sum_{i=1}^n \eta_{+, i} \l_i   \big) \, { \sum_{i<j} [ij]
\sum_{k<l}[kl] \over [ 12 ] \cdots [ n1 ]} \eta_1^4 ..
\eta_{i}^3 .. \eta_{j}^3..\eta_{k}^3 .. \eta_{l}^3 .. \eta_n^4 \nonumber\\
&=& { \sum_{i<j}
\langle ij \rangle [ij] \sum_{k<l} \langle kl \rangle [kl] \over [
12 ] \cdots [ n1 ]} \eta_1^4 \cdots \eta_n^4 \nonumber\\
&=&
 { q^4 \over [ 12 ]
\cdots [ n1 ]} \eta_1^4 \cdots \eta_n^4 ~.
\label{endcalc}
\eea
Equation \eqref{endcalc}
shows that there is a non-vanishing maximally non-MHV form factor for the self-dual field strength, whose expression is precisely given by \eqref{mnmhv}.

\subsection{Form factor of the complete stress-tensor multiplet }

In this section we consider the form factor of the the full, non-chiral
stress-tensor multiplet ${\cal T}(x, \theta^+, \tilde\theta_-)$.
We can write this as%
\footnote{Notice that the second equality is true only up to equations of motion because the non-chiral algebra closes only on shell. In the following we will work at tree level and hence this point will not affect our considerations.}
\bea
\label{nct}
\cT(x, \theta^+, \tilde\theta_-) & := & \Tr (W^{++} W^{++}) \nonumber \\
&=& e^{i \theta^+Q_+ +
i \tilde\theta_- \bar Q^- }\,  {\rm Tr}(\phi^{++}\phi^{++})(x) \, e^{-i \theta^+Q_+ -
i \tilde\theta_- \bar Q^- }\\
&=& {\rm Tr}(\phi^{++}\phi^{++}) +  (\theta^+)^4
{\cal L}+ (\tilde\theta_-)^4 \tilde{\cal L} + (\theta^+
\sigma^\mu \tilde\theta_-) (\theta^+ \sigma^\nu \tilde\theta_-)
T_{\mu\nu} + \cdots  ~, \nonumber
 \eea
where we have indicated only some terms of the full multiplet.

The right-hand side of \eqref{nct} is an expansion in the chiral as well as anti-chiral variables $\theta^+$ and $\tilde\theta_{-}$.
We can parallel this feature in the states by using a non-chiral description as in  \cite{Huang:2011um} with fermionic variables
$\eta_+$ and  $\tilde\eta^-$.   With this choice, the  supersymmetry algebra is realised on states as
\bea
&&\lan \, i \, |  Q_+  \, = \, \lan \, i \, | \lambda_i \eta_{+,i}
~, \qquad \lan \, i \, | Q_-  \, = \, \lan \, i \, | \lambda_i {\partial \over
\partial\tilde\eta^-_i} \, ,
\nonumber\\ &&  \lan \, i \, | \bar Q^- \, = \, \lan \, i \, | \tilde\lambda_i
\tilde\eta^-_i \, ,  \qquad \ \
\lan \, i |\, \bar Q^+
= \lan \, i \, | \tilde\lambda_i {\partial \over
\partial\eta_{+,i}} \, .
 \eea
This non-chiral representation can be obtained via a simple Fourier transform of half of the chiral superspace variables.
In terms of the Nair description of states, this amounts to introducing a
new super wavefunction,
\begin{eqnarray}
\Phi(p,\eta_+, \tilde\eta^-) &:=& \int\!\!d^2 \eta_- \
e^{i\eta_- \tilde\eta^-} \Phi(p,\eta)  \\&=& g^+(p) (\tilde\eta^-)^2 + \cdots +
\phi^{++} (\eta_+)^2 (\tilde\eta^-)^2 + \phi^{--} + \cdots + g^-(p)
(\eta_+)^2 ~. \qquad \nonumber
\end{eqnarray}
As a result, operators and superstates  live in a non-chiral superspace.
The non-chiral form factor in this representation is defined as
\beq
\label{ncff}
{\cal F}(q, \gamma_+, \tilde\gamma^-; 1, \ldots , n)  \ :=  \
\int\!\!d^4x\, d^4 \theta^{+}\, d^4 \tilde\theta_{-} \,
e^{-i( qx + \theta^+ \gamma_+ + \tilde\theta_- \tilde\gamma^-) } \,
\langle 1 \cdots n  |  \,  \cT(x, \theta^+, \tilde\theta_-)| 0  \ran \, .
\eeq
In order to write down Ward identities for \eqref{ncff}, we consider the action of supersymmetry generators on the operator ${\cal T}(x, \theta^+, \tilde\theta_-)$:
\bea
&& \hskip -.6cm [Q_+ , \cT(x, \theta^+, \tilde\theta_-) ]\,  =\,
i {\partial \over
\partial\theta^+} \cT(x, \theta^+, \tilde\theta_-) ~, \quad [Q_-, \cT(x, \theta^+, \tilde\theta_-) ] \, =\,
- \tilde\theta_- {\partial \over \partial x} \cT(x, \theta^+,
\tilde\theta_-) ~,
\nonumber\\
&& \hskip -.6cm [ \bar Q^- , \cT(x, \theta^+, \tilde\theta_-) ] \, =\,
-{\partial \over
\partial\tilde\theta_-} \cT(x, \theta^+, \tilde\theta_-) ~, \quad [ \bar Q^+, \cT(x, \theta^+, \tilde\theta_-) ]
\,  =\,
i \theta^+ {\partial \over \partial x} \cT(x, \theta^+,
\tilde\theta_-) ~. \nonumber\\
\eea
Following closely the derivation of  the Ward identities described in the
previous section, we arrive at the following relations for each
supersymmetry  generator,
\bea
\label{holiday}
Q_+ : && (\eta_+ \lambda - \gamma_+) {\cal F} = 0 ~, \qquad
Q_- : \quad \Big(q {\partial \over \partial \tilde\gamma^-} - \lambda
{\partial \over \partial \tilde\eta^-} \Big) {\cal F} = 0 ~, \\ \bar
Q^- :
&& (\tilde\eta^- \tilde\lambda - \tilde\gamma^-) {\cal F} = 0 ~, \qquad
\bar Q^+ : \quad \Big(q {\partial \over \partial \gamma_+} -
\tilde\lambda {\partial \over \partial \eta_+} \Big) {\cal F} = 0
~, \eea
and hence the  form factor in \eqref{ncff} takes the form
\bea
\label{ffs4}
{\cal F} = \delta^{(4)} ( q - \sum_{i=1}^n \l_i \lt_i) \, \delta^{(4)}\big( \gamma_{+} -  \sum_{i=1}^n
\eta_{+, i} \l_i   \big) \, \delta^{(4)}\big( \tilde\gamma^{-} -
\sum_{i=1}^n \tilde\eta_{i}^- \tilde\l_i   \big) \,
{\cal F}_{\phi^2}^{\rm nc} ~,
\eea
for some function ${\cal F}_{\phi^2}^{\rm nc}$.

A useful observation is that  ${\cal F}_{\phi^2}^{\rm nc} $  can be
obtained from the corresponding function introduced in \eqref{ecf} for the  chiral form factor  via a
half-Fourier transform on the  $\eta$ and $\tilde\eta$ variables, as
\bea
{\cal F}_{\phi^2}^{\rm nc}(\lambda, \tilde\lambda, \eta_+,
\tilde\eta^-)\  = \ \prod_{i=1}^n \int\!\!d^2 \eta_{-,i} \ e^{i \eta_{-,i}
\tilde \eta_i^-} \,  {\cal F}_{\phi^2}(\lambda,
\tilde\lambda, \eta_+, \eta_-) ~.
\eea
In the remaining part of this section we would like to  show a few applications of this formulation.

To begin with, we specialise to the MHV case, for which we have
\bea
{\cal F}_{\phi^2}^{\rm MHV, \, nc}
&=&
\prod_{i=1}^n \int\!\!d^2 \eta_{-,i} \
e^{i \eta_{-,i} \tilde \eta_i^-} \  { \delta^{(4)} \big(
\sum_{i=1}^n \eta_{-, i}  \l_{i} \big) \over \langle 12 \rangle
\cdots \langle n1 \rangle}
\nonumber\\ &=&
{\lan kl\ran^2 \over \lan12\ran \cdots
\lan n1 \ran} \prod_{i\neq k,l}^n (\tilde\eta_i^-)^2 + \cdots ~.
\eea
The MHV form factor of ${\rm Tr}(\phi^{+})^2$ is then obtained by  extracting  the
coefficient of $(\gamma_+)^4 (\tilde\gamma^-)^4$ in \eqref{ffs4}, and thus it is immediately seen to give the correct answer.
The form factor with an insertion of the chiral Lagrangian ${\cal L}$ (which includes ${\rm Tr}(F_{\rm SD}^2)$) is obtained by taking the coefficient of
$(\gamma_+)^0 (\tilde\gamma^-)^4$:
\bea {\cal F}_{\cal L}^{\rm MHV} \, =\,  \delta^{(4)}\big( \sum_{i=1}^n
\eta_{+, i} \l_i \big) {\cal F}_{\phi^2}^{\rm MHV}
\, = \,
 {\lan kl\ran^4 \over
\lan12\ran \cdots \lan n1 \ran} \Big( \eta_{+,k}^2 \eta_{+,l}^2
\prod_{i\neq k,l}^n (\tilde\eta_i^-)^2 \Big) + \cdots
\ , \eea
as expected.
Finally, in order to obtain the form factor with $\tilde{\cal L}$
(which includes ${\rm Tr}(F_{\rm ASD}^2)$),
we extract the coefficient of $(\gamma_+)^4 (\tilde\gamma^-)^0$:
\bea
{\cal F}_{\tilde{\cal L}}^{\rm MHV}
&=&
\delta^{(4)}\big(
\sum_{i=1}^n \tilde\eta_{i}^- \tilde\l_i \big) {\cal
F}_{\phi^2}^{\rm MHV} \ = \ {\sum_{i<j} \langle ij \rangle [ij] \sum_{k<l} \langle kl
\rangle [kl] \over \lan12\ran
\cdots \lan n1 \ran} \prod_{i=1}^n (\tilde\eta_i^-)^2
\nonumber\\
 &  = &   {q^4 \over
\lan12\ran \cdots \lan n1 \ran} \prod_{i=1}^n (\tilde\eta_i^-)^2
\, ,
\eea
which is indeed also correct.

\section{Supersymmetric methods \label{section-supermethod}}

In this section we take a brief survey of various methods that can be used to calculate form factors of the complete stress-tensor multiplet, at tree and loop level. These are simple but interesting extensions of well-known techniques for scattering amplitudes -- MHV diagrams \cite{csw}, on-shell recursion relations \cite{bcf, bcfw} and (generalised) unitarity \cite{bddk, fusing,bdkgen,bcfgen} --  thus we will limit ourselves to highlighting the peculiarities we encounter when dealing with  form factors. The non-supersymmetric versions of these methods have been considered earlier in Section 2 and in \cite{bsty2}.

A preliminary observation is that the form factor of the complete
stress-tensor multiplet operator  $\cT(x, \theta^+, \tilde\theta_-)$  can be  expressed in terms of that  of its lowest bosonic component  $\Tr (\phi^{++} \phi^{++})$,  as we have shown in  \eqref{ffs4}, namely
\bea
 {\cal F} =\delta^{(4)} ( q - \sum_{i=1}^n \l_i \lt_i )
\,
\delta^{(4)}\big( \gamma_{+} - \sum_{i=1}^n \eta_{+,
i} \l_i   \big) \, \delta^{(4)}\big( \tilde\gamma^{-} - \sum_{i=1}^n
\tilde\eta_{i}^- \tilde\l_i   \big) \, {\cal F}_{\phi^2}^{\rm nc} ~,
 \eea
where $\cF_{\phi^2}^{\rm nc}  := \lan 1 \cdots n  | \Tr ( \phi^{++} \phi^{++}) (0) | 0 \ran$ and the superstate $\lan 1\cdots n |$ is here in the non-chiral representation. One can then switch instantly to the chiral representation via a half-Fourier transform from
the $\tilde\eta^{-}$ to the $\eta_{+}$ variables. Hence, we only need to devise methods to calculate the form factor
 $\lan 1 \cdots n  | \Tr ( \phi^{++} \phi^{++} )(0) | 0 \ran$ using a chiral representation for the external state. This is the problem we address in the following.%
 \footnote{To simplify our notation, we will drop from now on the subscript in
 ${\cal F}_{\phi^2}$.}

\subsection{Supersymmetric  MHV rules \label{section-superMHV}}

\begin{figure}[h]
\centerline{\includegraphics[height=2.7cm]{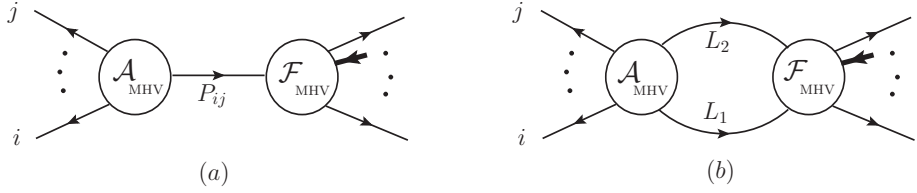} }
\caption{\it {\rm (a)} MHV diagram for a tree-level NMHV form factor. {\rm (b)}
MHV diagram for a one-loop MHV form factor.} \label{SusyFF-MHVrule}
\end{figure}

We begin with a lightning illustration of super MHV rules. Here, the super MHV  form factor,
\bea
 {\cal F}^{\rm MHV}
 (1,2,\cdots, n;q) = {\delta^{(4)} (q - \sum_{i} \l_i \lt_i)
\, \delta^{(4)}(\sum_i \lambda_i
\eta_{i,-}) \over \langle 1~2 \rangle \langle 2~3 \rangle \cdots
\langle n~1 \rangle} ~,
\eea
is continued off shell with the standard prescription \eqref{off1} of  \cite{csw}, and used as a vertex in addition to the standard MHV vertices.  Form factors have a single operator insertion, hence we only draw diagrams with a single form factor MHV vertex. As an example, consider the NMHV tree-level super form factor. It can be computed by summing over all diagrams in Figure \ref{SusyFF-MHVrule}$(a)$, whose expression is
\bea
\hspace{-.7cm}{\cal F}^{(0)}_{\rm NMHV} \!\!\!\!&\!\!\!=\!\!\!&\!\!\!\! \sum_{i=1}^n \sum_{j=i+1}^{i+n-2}
\int\!\!d^4 P_{ij} \,\int\!\!d^4\eta_P ~ {\cal A}^{(0)}_{\rm MHV}( i, .. , j,
P_{ij}) ~ {1\over P_{ij}^2} ~ {\cal F}^{(0)}_{\rm MHV}(j\!+\!1,
.., i\!-\!1, -\!P_{ij}; q)
\nonumber\\
\hspace{-.3cm}\!\!\!\!&=&\!\!\!\! {\cal F}^{(0)}_{\rm MHV}
\sum_{i=1}^n \sum_{j=i+1}^{i+n-2} {\langle i\!-\!1~i \rangle \langle
j~j\!+\!1 \rangle \over \langle i\!-\!1~P_{ij} \rangle
\langle P_{ij}~i \rangle \langle j~P_{ij} \rangle\langle P_{ij}~j\!+\!1
\rangle} {1\over P_{ij}^2} \delta^{(4)}\Big( \sum_{k=i}^{j} \langle
P_{ij}~k\rangle \eta_k^A \Big) .
\eea
We have also calculated tree-level {$\mathrm{N}^2$MHV} super form factor up to
six points  and checked that the results are all independent of the choice of
reference spinor.  We have also re-derived the split-helicity form factors, and checked numerical  agreement with the results presented in Section \ref{section-splithelicity}.

As an additional example, consider the one-loop MHV super form factor. Following \cite{bst}, this can be computed by summing over all diagrams in Figure \ref{SusyFF-MHVrule}$(b)$, and is given by
\beqa
\ {\cal F}^{(1)}_{\rm MHV} &=& \sum_{i=1}^n
\sum_{j=i}^{i+n-1} \int\!\!{d^D L_1\over L_1^2 + i \varepsilon} \,{ d^D L_2\over L_2^2 + i \varepsilon}  \,
 \int\!\!d^4 \eta_{L_1} \int\!\!d^4 \eta_{L_2} \\
&& {\cal A}^{(0)}_{\rm MHV} \big( i \,  \ldots, j, L_1 , L_2 \big)\,
{\cal F}^{(0)}_{\rm MHV} (-L_2 , - L_1, j\!+\!1,  \ldots, i\!-\!1;
q) \ . \nonumber
\eeqa
%
%
%
%
%
%
%

Finally, we note that the MHV vertex expansion may be proved at tree level along the lines of
\cite{Elvang:2008vz},  namely by using a BCFW recursion relation with an all-line shift and showing that
this is identical to the MHV diagram expansion.

\subsection{Supersymmetric  recursion relations}

\begin{figure}[h]
\centerline{\includegraphics[height=2.4cm]{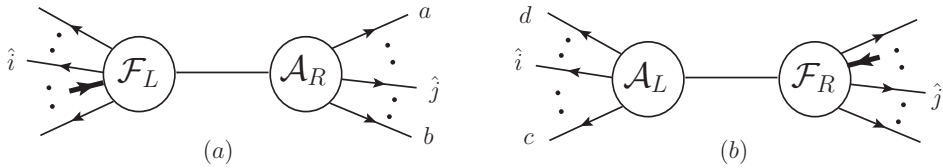} }
\caption{\it The two recursive diagrams discussed in the text.}
\label{SusyFF-BCFW}
\end{figure}

Now we consider a simple extension of the supersymmetric version \cite{bhtdual,ahck}
of the BCFW recursion relation \cite{bcf, bcfw}.
We choose to work with an $[i,j\rangle$ shift,
$\tilde\lambda_i \rightarrow \tilde\lambda_i + z \tilde\lambda_j$,
$\lambda_j \rightarrow \lambda_j - z \lambda_i$, $\eta_i \rightarrow
\eta_i + z \eta_j$. Factorisation requires that each term in the recursion relation must contain one form factor and one amplitude.
Hence,  for each kinematic channel we need to sum over two
diagrams, with the form factor appearing either on the
left-hand or right-hand side, see Figure \ref{SusyFF-BCFW}.
The result one obtains by summing over these two classes of
diagrams has the form
\bea
 {\cal F}(0) &=& \sum_{a,b} \int d^4 P d^4 \eta_P \, {\cal
F}_L(z\!=\!z_{ab}) {1\over P_{ab}^2} {\cal A}_R(z\!=\!z_{ab})
\nonumber\\ & + & \sum_{c,d} \int d^4 P d^4 \eta_P \, {\cal
A}_L(z\!=\!z_{cd}) {1\over P_{cd}^2} {\cal F}_R(z\!=\!z_{cd}) \ .
\eea
%
%
%
%
%
%
%
One point deserves a special attention, namely the large-$z$ behaviour of the form factor. Recall that in order to have a recursion relation without boundary terms we need $\cF (\ldots \hat{p}_i, \ldots , \hat{p}_j , \ldots ) \to 0$ as $z\to \infty$. We discuss this important point in
Appendix \ref{appendix-largez}, where we prove that the condition mentioned above is indeed satisfied. We would also like to point out that the basic seeds in the form factor recursion relation are the two-point form factor, together with the  three-point amplitudes.

\subsection{Supersymmetric  unitarity-based method}

\begin{figure}[h]
\centerline{\includegraphics[height=2.6cm]{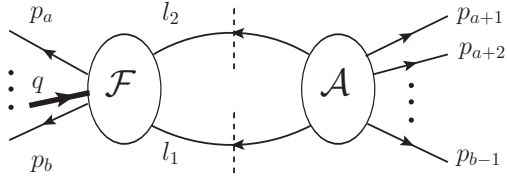} }
\caption{\it A two-particle cut diagram for a one-loop form factor. }
\label{SusyFF-Unitarity}
\end{figure}

Supersymmetric generalised unitarity, as well as supersymmetric MHV rules, are easily applied to form factors. Consider for example a two-particle cut, depicted in Figure \ref{SusyFF-Unitarity}. On one side of the cut we have a tree-level form factor, on the other a tree scattering amplitude.
For the case of a one-loop supersymmetric MHV form factor, the two-particle cut is equal to
\beqa
\left. {}  \ {\cal F}^{(1)}_{\rm MHV} \right|_{s_{a+1, b-1
}\!-\!\mathrm{cut}} &=& \int\!\!d{\rm LIPS}(l_1, l_2; P) \int\!\!d^4 \eta_{l_1} \int\!\!d^4 \eta_{l_2} \\
&& {\cal F}^{(0)}_{\rm MHV} (-l_2 , - l_1, b,  \ldots, a; q) {\cal
A}^{(0)}_{\rm MHV} \big(l_1 , l_2, (a+1) \,  \ldots, (b-1) \big) \ ,
\nonumber
\eeqa
where the Lorentz-invariant phase-space measure is
\beq d{\rm LIPS}(l_1, l_2; P) := d^D l_1\, d^D l_2 \, \delta^+
(l_1^2) \delta^+ (l_2^2) \delta^D (l_1+ l_2  + P)\ .
\eeq
%
%
%
%
%
%
The sum over all possible states which can propagate in the loop is automatically performed by the
fermionic integration.  A simple calculation gives
\beq \left. {}  \
 {\cal F}^{(1)}_{\rm MHV}
\right|_{s_{a+1, b-1 }\!-\!\mathrm{cut}} = {\cal F}^{(0)}_{\rm MHV}
\int\!\!d{\rm LIPS} (l_1, l_2; P_{a+1,  b-1} ) \ {\lan a \, a+1 \ran
\lan l_2\, l_1 \ran \over \lan a \, l_2 \ran \lan l_2 \, a+1 \ran }
\ {\lan b-1 \, b \ran \lan l_1\,  l_2 \ran \over \lan b-1 \, l_1
\ran \lan l_1 \, b\ran } \ ,
\eeq
which reproduces the result  derived in \cite{bsty2} using component form factors and amplitudes.

\section{Dual MHV rules for form factors}

It was shown in  \cite{Mason:2010yk} that the expectation value of supersymmetric Wilson loops in momentum twistor space generates all planar amplitudes in $\cN=4$ SYM, and  dual MHV rules in momentum twistor space  were proposed in \cite{Bullimore:2010pj}.
Inspired by these results, dual MHV rules directly formulated in dual momentum space were introduced in   \cite{bsty1}.
In these rules a
lightlike closed polygon formed by linking the
on-shell momenta of the external particles following their colour ordering plays an important role. Note that the same polygon appears in the amplitude/Wilson loop duality \cite{am,dks,bht}.

In this section we extend these rules to the calculation of form factors of the special operator considered in previous sections, namely the chiral part of the stress-tensor multiplet operator. It turns out that the rules for the amplitude have to be modified only slightly. More precisely, there are no new vertices to be introduced, and we only have to modify (super)momentum conservation of the particles in order to account for the (super)momentum injected by the operator. In the dual momentum picture, this implies the breaking of the closed null contour describing the particle's momenta.

The vertices of this open polygon in dual supermomentum space are labelled by $(x_i, \Theta_i)$  \cite{dhks}, with%
\footnote{In order to avoid confusion with the variables $\theta$'s introduced in earlier sections, we denote by  $\Theta$ the variables living in dual super momentum space.}
\bea
x_i - x_{i+1} := p_i = \lambda_i \tilde \lambda_i ~, \qquad
\Theta_i - \Theta_{i+1} := \lambda_i \eta_i ~,
\eea
with
\bea
x_i - x_{i+n} = \sum_{j=1}^n p_j = q ~, \qquad
\Theta_{i} -
\Theta_{i+n} = \sum_{j=1}^n \lambda_j\eta_{j} = \gamma ~,
\eea
where  $q$ ($\gamma$) is the (super)momentum carried by the operator.
Note that in the previous equation we have effectively injected the (super)momentum of the operator between
on-shell states labelled by $i-1$ and $i$ and this is where the breaking of the polygon occurs.
For each diagram an appropriate choice for the location of the breaking will have to be made. Furthermore,
in this section  we  consider the chiral operator $\cT(x, \theta^+)$ for which  $\gamma_{-}=0$, and hence $\Theta_{i; - } - \Theta_{i+n; - }=0$.
For amplitudes we have of course $q=0$ and $\gamma=0$ which would bring us back to a closed
lightlike polygon.

\begin{figure}[h]
\centerline{\includegraphics[height=2.7cm]{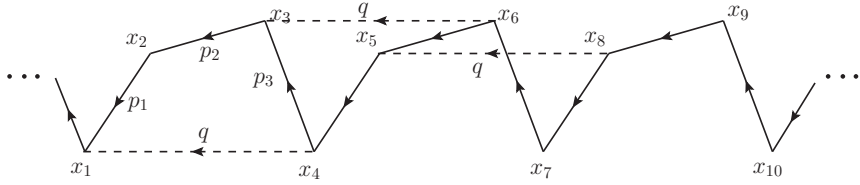} }
\caption{\it The kinematic configuration in dual momentum space used to calculate three-point form factors using dual MHV rules. }
\label{periodicWL-ST}
\end{figure}

In practice it is useful to convert the open polygon for form factors into a periodic configuration in
dual momentum space with period $q$ ($\gamma$) in the bosonic (fermionic) direction as in
Figure \ref{periodicWL-ST}.
This is partially motivated by a duality observed at strong coupling in \cite{Alday:2007he, mz} where form factors are related to the area of minimal surfaces ending on an infinite periodic sequence of null segments at the boundary of $AdS$. In \cite{bsty2} an attempt  was made to map this geometric picture to weak coupling, in a way similar to the amplitude/Wilson loop duality \cite{dks,bht}.

The emergence of a periodic configuration is also natural from a field-theoretic point of view once one takes into account that the operator insertion is a colour singlet, and hence does not interfere with the colour ordering of the external state. In other words, the (super)momentum carried by the operator can be inserted between any pair of particle momenta without spoiling the ordering.  Precisely by resorting to a periodic configuration we can account for this property, as Figure \ref{periodicWL-ST} clearly shows.

\begin{figure}[h]
\centerline{\includegraphics[height=2.2cm]{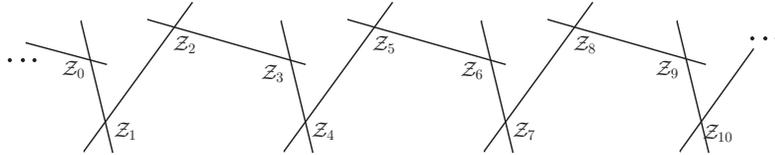} }
\caption{\it The same kinematic configuration presented in Figure \ref{periodicWL-ST}, in terms of momentum twistor space variables.  } \label{periodicWL-MT}
\end{figure}

One can also consider this periodic kinematic configuration in momentum twistor space \cite{hodges}, as shown
in Figure \ref{periodicWL-MT}, with space-time points being mapped to  lines in twistor space:
$(x_i,\Theta_i) \sim {\cal Z}_{i-1} \wedge {\cal Z}_i$,  where
\bea
{\cal Z}_i = (\lambda_i, \nu_i, \chi_i) ~, \qquad \nu_i = x_i
\lambda_i = x_{i+1} \lambda_i ~, \qquad \chi_i = \Theta_i \lambda_i
= \Theta_{i+1} \lambda_i ~.
\eea
%

%
%

\subsection{Examples}

In this section we want to explain the  dual MHV  rules  by discussing a number of simple examples of tree-level and one-loop form factors.
The dual MHV rules in dual momentum space for $\cN=4$ amplitudes are summarised for the reader's convenience  in Appendix \ref{appendix-dualMHV}, and we refer to \cite{bsty1} for full details.

\begin{figure}[h]
\centerline{\includegraphics[height=2.7cm]{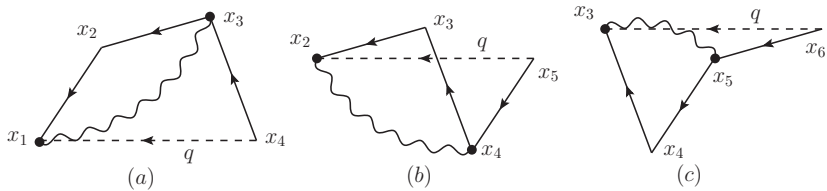} }
\caption{\it Dual MHV diagrams for the three-point tree NMHV form
factor.} \label{DualMHV-tree}
\end{figure}

\begin{figure}[h]
\centerline{\includegraphics[height=2.3cm]{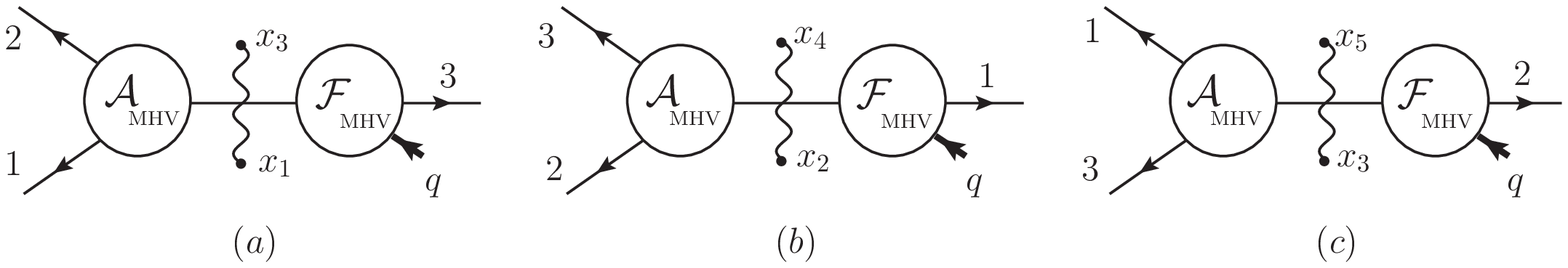} }
\caption{\it Corresponding MHV diagrams for the three-point tree NMHV form
factor.} \label{MHV-tree}
\end{figure}

The first example  is that of an  NMHV three-point form factor. The corresponding diagrams are shown  in Figure
\ref{DualMHV-tree}, and are in one-to-one correspondence  with  three conventional MHV diagrams, depicted in Figure \ref{MHV-tree}. Notice that the three diagrams in Figure \ref{DualMHV-tree} can be obtained by selecting the appropriate period in Figure \ref{periodicWL-ST}.


The extension to  $n$-point NMHV form factors is immediate --  we consider all  dual MHV diagrams where one propagator connects
two external vertices {\it within one period.} The final result is given by summing over all translationally inequivalent diagrams as
\beq
\label{where}
 {\cal F}_{\rm NMHV}^{(0)} = {\cal F}_{\rm MHV}^{(0)}
\sum_{i=1}^n \sum_{j=i+2}^{i+n-1} {\langle i\!-\!1 ~i \rangle \over
\langle i\!-\!1 ~\ell_{ij} \rangle \langle \ell_{ij} ~i \rangle}
{\langle j\!-\!1 ~j \rangle \over \langle j\!-\!1 ~\ell_{ij} \rangle
\langle \ell_{ij} ~j \rangle} {1\over x_{ij}^2} \int d^4 \eta_{ij} ~
\delta^{0|8} (\ell_{ij} \eta_{ij} + \Theta_{ij}) ~,
\eeq
where the spinor $|\ell_{ij}\rangle$ associated to the internal leg is defined as
\bea
| \ell_{ij} \rangle :=  |x_{ij}|\xi ] ~,
\eea
and where $|\xi]$ is an arbitrary reference spinor.  Notice that the particle labels of spinor variables $i$ and $i+n$  are identified in this expression.
Importantly, the fact that we are calculating a form factor rather than an amplitude --  and the
corresponding dependence on $q$ and $\gamma$ -- is completely
encoded in the periodic kinematic configuration as defined earlier.
Furthermore, we observe  that
every diagram in the sum corresponds to a particular period (see Figures \ref{DualMHV-tree} and \ref{MHV-tree}).

Notice that diagrams where a propagator connects two adjacent points give a vanishing result, and  therefore are not included in the summation.
On the other hand, diagrams where a propagator connects two points separated by  exactly one period or more are non-vanishing, and have to be  excluded since there is no corresponding conventional MHV diagram.
For instance, among the  three-point diagrams in Figure \ref{DualMHV-tree} we do not include the diagram with a propagator connecting points $x_1$ and $x_4$.
This is an example of a more general fact:
\textit{diagrams where a single propagator connects points $x_i$ and $x_j$ with $|i-j| \geq n$ have to be discarded. } This applies to any loop order.
The reason for this rule is that there are no corresponding supersymmetric MHV diagrams.

As an aside we mention that \eqref{where} can also be written in terms of momentum twistor variables as
\bea {\cal F}_{\rm NMHV}^{(0)} = {\cal F}_{\rm MHV}^{(0)}\,
\sum_{i=1}^n \sum_{j=i+2}^{i+n-1} [*, i\!-\!1, i, j\!-\!1, j] ~,
\eea
where ${\cal Z}_*$ is the reference momentum twistor, chosen as
\bea {\cal Z}_* = (0, \xi, 0) ~,
\eea
and $[*, i\!-\!1, i, j\!-\!1, j]$ is defined in \eqref{R-inv}.

\begin{figure}[h]
\centerline{\includegraphics[height=3cm]{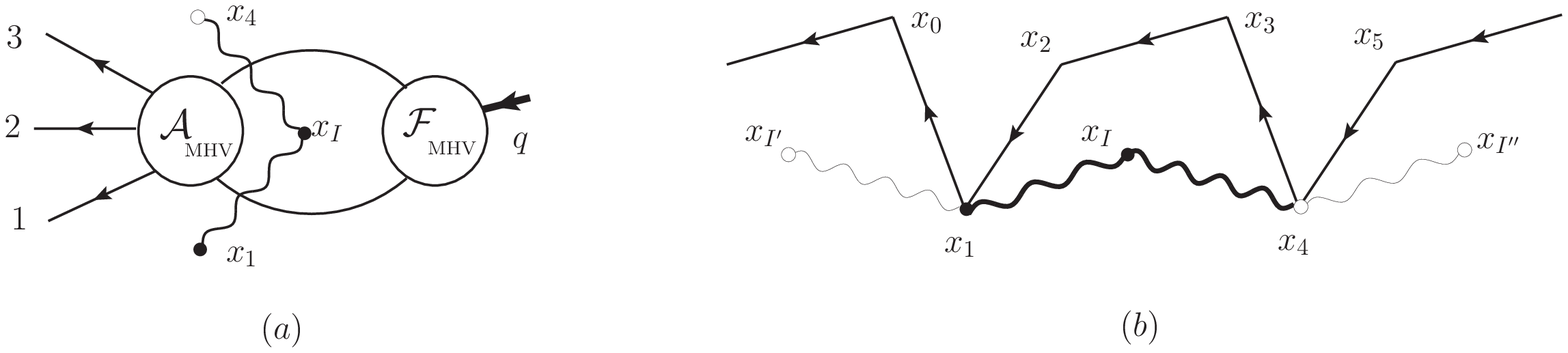} }
\caption{\it
A special diagram with   two propagators with  momenta  $x_{iI}$ and $x_{i\!+\!n I}$.
In the dual MHV diagram there are two propagators with momenta $x_{1I}$ and $x_{4I}$,
 and two vertices, $x_{1}$ and $x_I$.
Such diagrams correspond to the last three lines of \eqref{ddd}.}
\label{DualMHV-S1loop}
\end{figure}

The case of one-loop MHV form factors is similar to the tree-level NMHV case. The $n$-point one-loop MHV form factor is given by
\bea
\label{ddd}
{\cal F}_{\rm MHV}^{(1)} &=&  {\cal F}_{\rm MHV}^{(0)}\, \int\!d^4
x_I d^8 \Theta_I \, \sum_{i=1}^n \sum_{j=i+1}^{i+n-1}  \ {\langle i\!-\!1
~i \rangle \over \langle i\!-\!1 ~\ell_{iI} \rangle \langle
\ell_{iI} ~i \rangle} {\langle j\!-\!1 ~j \rangle \over \langle
j\!-\!1 ~\ell_{Ij} \rangle \langle \ell_{Ij} ~j
\rangle}  \\
&& {1\over \langle \ell_{iI} \ell_{Ij} \rangle \langle
\ell_{Ij} \ell_{iI} \rangle } ~ {1\over x_{iI}^2} \int d^4 \eta_{iI}
~ \delta^{0|8} (\ell_{iI} \eta_{iI} + \Theta_{iI}) {1\over x_{Ij}^2}
\int\!d^4 \eta_{Ij} ~ \delta^{0|8} (\ell_{Ij} \eta_{Ij}
+
\Theta_{Ij}) \nonumber\\  & + &
 {\cal F}_{\rm MHV}^{(0)}\, \int\!\!d^4 x_I d^8 \Theta_I \,
\sum_{i=1}^n \ {\langle i\!-\!1
~i \rangle \over \langle i\!-\!1 ~\ell_{iI'} \rangle \langle \ell_{iI'} ~\ell_{iI} \rangle \langle
\ell_{iI} ~i \rangle} ~ {1\over \langle \ell_{iI} \ell_{i\!+\!n I} \rangle \langle
\ell_{i\!+\!n I} \ell_{iI} \rangle } \nonumber \\
&& {1\over x_{iI}^2} \int\!d^4 \eta_{iI}
~ \delta^{0|8} (\ell_{iI} \eta_{iI} + \Theta_{iI}) {1\over x_{i\!+\!n I}^2}
\int\!d^4 \eta_{i\!+\!n I} ~ \delta^{0|8} (\ell_{i\!+\!n I} \eta_{i\!+\!n I} +
\Theta_{i\!+\!n I}) \nonumber\\ && \int\!\!d^4 x_{I'} d^8 \Theta_{I'} \delta^4(x_{I'I}
- x_{i i\!+\!n}) \delta^{0|8}(\Theta_{I'I} - \Theta_{i i\!+\!n}) ~.
\nonumber \eea
Notice that we have treated a special class of diagrams differently, corresponding to the last three  lines in \eqref{ddd}.
These are diagrams where the two propagators have momenta  $x_{iI}$ and $x_{i\!+\!n I}$.
An example of such a  diagram in the case of a three-point form factor is shown in Figure \ref{DualMHV-S1loop}.

\begin{figure}[h]
\centerline{\includegraphics[height=7cm]{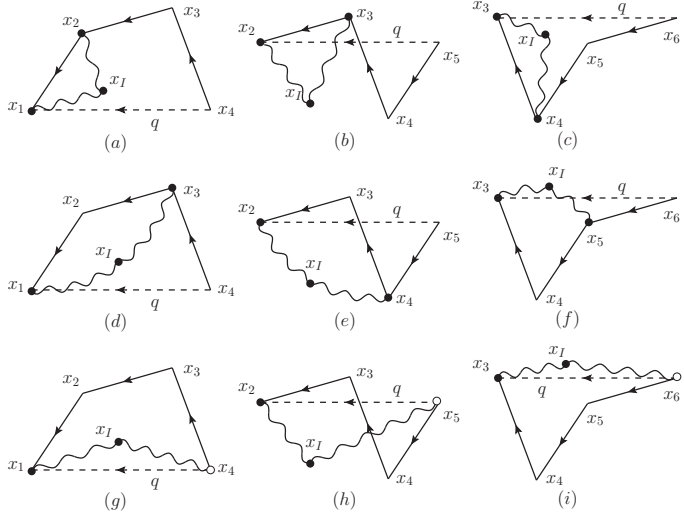} }
\caption{\it Dual MHV diagrams for the three-point MHV form
factor at one loop.} \label{DualMHV-oneloop}
\end{figure}

The three-point  dual MHV diagrams at one loop  are shown in Figure \ref{DualMHV-oneloop}.
The diagrams in Figure \ref{DualMHV-oneloop}  (g)-(i) are of the special class described earlier in  Figure \ref{DualMHV-S1loop}.
Note that in the case of loop diagrams we also have to include diagrams
where two adjacent
points or two points separated by exactly one period are connected by two or more propagators
(see Figure \ref{DualMHV-oneloop} diagrams (a)-(c) and (g)-(i) respectively).
We should also stress that all diagrams where two points $x_i$ and $x_j$ with $|i-j|>n$ are
connected must be discarded.
Generalisations to non-MHV form factors are straightforward.

Finally we  compare the dual MHV diagrams with the periodic Wilson line diagrams studied in \cite{bsty2}.
We can see that an identical truncation was necessary in order to obtain the correct result:
in a single MHV diagram the external vertices which are connected to propagators
must reside  within one period, and the whole form factor is obtained by summing over all translationally inequivalent diagrams.

\subsection{Higher-loop diagrams}

At higher loops, the situation becomes more involved. To illustrate the main novelty we consider  the two-point MHV form factor at two loops.%
\footnote{Incidentally, we recall that while at one loop it has been proved that (four-dimensional) MHV diagrams reproduce complete amplitudes \cite{ftt},
there is no such statement at two loops and beyond. However, MHV diagrams at two loops and beyond can be used effectively to compute unregulated integrands of amplitudes (and form factors, as demonstrated here) which have recently attracted great interest in their own right \cite{nima}.
}

\begin{figure}[h]
\centerline{\includegraphics[height=3.4cm]{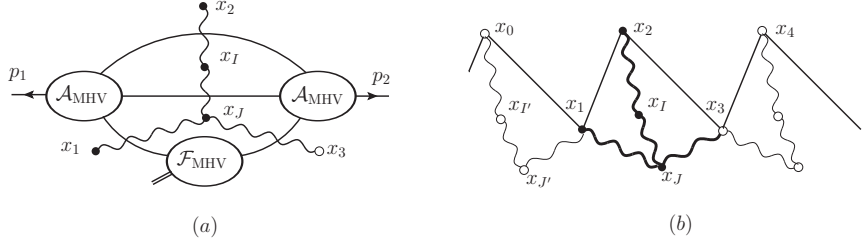} }
\caption{\it (a) First  MHV diagram for a two-loop, two-point  MHV
form factor. (b) The corresponding dual MHV diagram.  } \label{MHV2loop-I}
\end{figure}

\begin{figure}[h]
\centerline{\includegraphics[height=3.6cm]{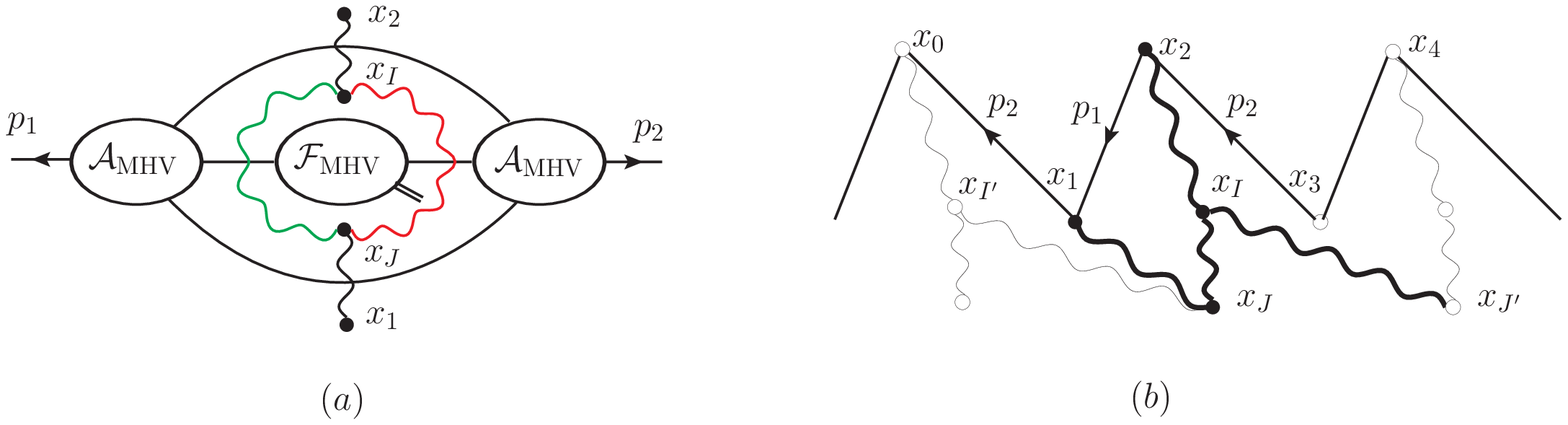} } \caption{\it
(a) Second  MHV diagram for   a two-loop, two-point  MHV form
factor. (b) The corresponding dual MHV diagram.} \label{MHV2loop}
\end{figure}

As prototypical examples, we consider  two particular diagrams,  depicted in Figures \ref{MHV2loop-I} and \ref{MHV2loop}. In the first diagram, the form factor MHV vertex is inserted in the exterior part of the diagram, whereas in the second situation it is inserted in the interior. On the right-hand side of each figure we also draw the corresponding dual MHV diagram. Let us start with the first, simpler situation. There is no subtlety in defining the internal region momenta $x_I$ and $x_J$. The momenta in the propagators  in the outer loop are $x_{2I}$, $x_{3J}$ and $x_{1J}$, and it is straightforward to write down the two-loop dual MHV integrand. In the notation of Appendix B, there are two internal vertices, two external  vertices at $x_1$ and $x_2$ (with $x_1$ being a two-point vertex) and four propagators, as shown by
dark bullets and dark wavy lines in  Figure \ref{MHV2loop-I} (b).

%
%

Consider now the second, more subtle situation drawn in Figure \ref{MHV2loop}.
In order to assign region momenta consistently to all regions in this diagram, we need to introduce
an additional loop momentum $x_{J'}$ such that $x_{J} - x_{J'} = q$, in exactly the same way as $x_1 - x_3 = q$.
Similarly, one can also introduce $x_{I'}$ such that $x_{I'} - x_{I} = q$.
The dual MHV diagram is  shown  in Figure \ref{MHV2loop}(b).

As before, we consider only translationally
inequivalent diagrams within one period. Each such diagram will have two one-point external vertices, two three-point internal vertices and four propagators, as shown by
dark bullets and dark wavy lines in Figure \ref{MHV2loop}(b).
The expression of this dual MHV diagram is then
\bea
\label{long}
&&
 \int\!\!d^4 x_I d^8
\Theta_I \
{1\over \langle \ell_{I2} \ell_{IJ} \rangle \langle \ell_{IJ}
\ell_{IJ'} \rangle \langle \ell_{IJ'} \ell_{I2} \rangle }
\int\!\!d^4 x_J d^8 \Theta_J \ {1\over \langle \ell_{J1} \ell_{JI'}
\rangle \langle \ell_{JI'} \ell_{JI} \rangle \langle \ell_{JI}
\ell_{J1} \rangle}
\nonumber \\
&& {\langle 1 2 \rangle \over \langle 1 \ell_{2I} \rangle \langle
\ell_{2I} 2 \rangle} {\langle 21 \rangle \over \langle 2 \ell_{1J}
\rangle \langle \ell_{1J} 1 \rangle}
\\ &&
 {1\over x_{I2}^2} \int\!\!d^4 \eta_{I2} ~ \delta^{0|8}
(\ell_{I2} \eta_{I2} + \Theta_{I2})  {1\over x_{J1}^2} \int\!\!d^4
\eta_{J1} ~ \delta^{0|8} (\ell_{J1} \eta_{J1} + \Theta_{J1})
\nonumber\\ && {1\over x_{IJ}^2} \int\!\!d^4 \eta_{IJ} ~
\delta^{0|8} (\ell_{IJ} \eta_{IJ} + \Theta_{IJ})  {1\over x_{IJ'}^2}
\int\!\!d^4 \eta_{IJ'} ~ \delta^{0|8} (\ell_{IJ'} \eta_{IJ'} +
\Theta_{IJ'})
\nonumber\\
&& \hskip -.5cm \int\!\!d^4 x_{I'} d^8 \Theta_{I'} \delta^4(x_{II'}
+x_{13}) \delta^{0|8}(\Theta_{II'} + \Theta_{13}) \int\!\!d^4 x_{J'}
d^8 \Theta_{J'} \delta^4(x_{JJ'} - x_{13}) \delta^{0|8}(\Theta_{JJ'}
- \Theta_{13}) ~. \nonumber \eea
Notice in the last line of \eqref{long}
the delta functions which enforce the periodicity of the super region momenta $x_{I'}$ and $x_{J'}$.
One can check that   \eqref{long}  is
indeed equivalent to the result of the conventional MHV diagram in
Figure \ref{MHV2loop}(a).

\begin{figure}[h]
\centerline{\includegraphics[height=3.5cm]{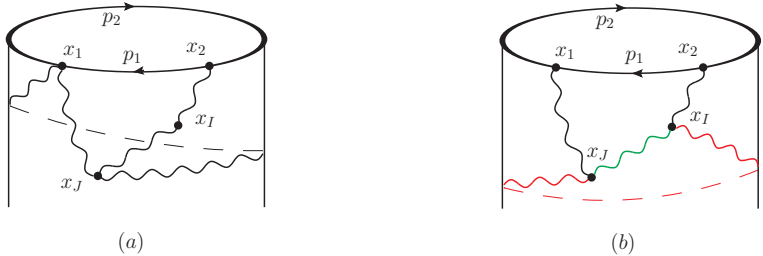} }
\caption{\it (a) Cylinder picture for the MHV diagram in Figure
\ref{MHV2loop-I}. (b) Cylinder picture for the MHV diagram in Figure
\ref{MHV2loop}. The period of the cylinder is $q$. }
\label{FFcylinder-2loop}
\end{figure}
The dual MHV rules for form factors described above can be understood more naturally if we put the periodic configuration on a cylinder of period $q$, see Figure \ref{FFcylinder-2loop}.
In particular, Figure \ref{FFcylinder-2loop}(b) corresponds to the  MHV diagram in Figure \ref{MHV2loop}.
The two coloured propagators connecting $x_I$ and $x_J$ form a loop with winding momentum $q$, which exactly
correspond to the coloured lines
in the  MHV diagram in Figure \ref{MHV2loop}(a).
We would like to stress a general feature of the rules we have described before, namely that no {\it single} propagator can stretch for one or more than one period around the cylinder.


The dual MHV rules can be applied to generic form factors.
As in the case of amplitudes, in order to
calculate an ${{\rm N}^k {\rm MHV}}$ form factor at $L$ loops,  we
need to sum over all allowed diagrams with
\bea \#({\rm internal \ vertices}) = L ~, \qquad \#({\rm
propagators}) = k + 2 L ~.
\eea
It would be very interesting to map the dual MHV rules described here to a dual  Wilson line picture for form factors.
We leave this question for future work.


\vspace{0.7cm}
\section*{Acknowledgements}

It is a pleasure to thank Fernando Alday, Nima Arkani-Hamed, Mathew Bullimore,
Bo Feng, Valeria Gili, Valya Khoze, Koji Hashimoto, Lionel Mason, Sri K.~Pattabhi Jois,
David Skinner, Wei Song, Bill Spence, Chung-I Tan,
Alexander Zhiboedov, and especially Rutger Boels and
Paul Heslop for sharing their insights. AB and GT would like to
thank the KITP at the University of California, Santa Barbara, where
their research was supported by the National Science Foundation
under Grant No.~NSF PHY05-51164, and
Nima Arkani-Hamed for an invitation to the Institute for Advanced Study,
Princeton, where some of the results of this paper were presented.
GY would like to thank the hospitality of Hamburg University, Harvard University and
Institute of Theoretical Physics, Chinese Academy of Sciences
where part of this work was done, and Paul Heslop
for an invitation to the $11^\mathrm{th}$ Workshop on Non-Perturbative
Quantum Chromodynamics, where the results of this paper were presented.
This work was supported by the STFC under a Rolling Grant ST/G000565/1.

\appendix

\section{Vanishing of form factors at large $z$ \label{appendix-largez}}

\subsection{Bosonic form factors \label{appendix-largez-bosonic}}

In this appendix we consider a generic non-MHV bosonic form factor of the operator ${\rm Tr}(\phi^2)$ and prove that, for a  $[k,l\rangle$ shift
\beq
\hat{\tilde{\lambda}}_k := \tilde{\lambda}_k + z \tilde{\lambda}_l \ , \qquad \qquad \hat\lambda_l := \lambda_l - z \lambda_k
\ ,
\eeq
$F(z)$ vanishes as  $z\rightarrow \infty$ if
\bea (h_k, h_l) \textrm{~ is equal to : ~} (0, +), ~(+, +), ~(-, +), ~(0, 0), ~(-, 0), ~(-, -)~. \eea
The proof is based on  the MHV diagram expansion of form factors,  and follows closely that for amplitudes presented   in \cite{bcfw}.

To begin with, it is immediate to see that an  MHV form factor \eqref{ffmhvtree} with a $[k,l\rangle$ shift vanishes as $z\rightarrow \infty$,
with the only exception of the case  $(h_k, h_l) = (+,0)$.
Consider now a generic  non-MHV form factor.
Each  MHV diagram contributing to its expansion is a product of MHV vertices, times propagators ${1/ L^2}$.
These propagators will  either be independent of $z$, or vanish when $z\rightarrow \infty$.
As in \cite{bcfw}, the spinors $\lambda_L = L|\tilde\xi ]$ associated to  internal legs  can also be made $z$-independent
by choosing the reference spinor $\tilde\xi$ to be equal to  $\tilde\xi = \tilde\lambda_l$.
Thus, dangerous $z$-dependent terms can only arise from terms affected by the shifts in the  external legs $k$ and $l$.

For the cases  where $(h_k, h_l)$ is $(\pm,+)$ or $(0,+)$, only the denominators acquire  $z$-dependence, and hence $F(z)$ vanishes at large $z$.  By  using anti-MHV diagrams we arrive at the same result for the case where $(h_k, h_l)$ is equal to either $(-,-)$ or $(-,0)$.
The case $(h_k, h_l) = (0,0)$ needs special attention. The case when $k$ and $l$ belong to the same MHV vertex has already been considered, and leads to a falloff of the diagram as $z\to \infty$.
When $k$ and $l$ belong to different vertices, there will be at least one propagator depending on $z$, which will provide a factor of $1/z$  at large $z$.   The vertex involving leg $l$ behaves asymptotically as
 $z^2 / z^2$  regardless of whether it is an MHV form factor or a conventional  MHV vertex,
 while all other vertices are independent of $z$. We conclude that each MHV diagram falls off as $1/z$ at large $z$.

We mention here that the argument described above can also been applied to scattering amplitudes.
Shifting two scalars makes the amplitude vanish as $z \rightarrow \infty$
provided that the scalars take the same $SU(4)$ indices.

\subsection{Supersymmetric form factors \label{appendix-largez-super}}

As we have shown in the previous appendix, the bosonic form factor vanishes at infinity for an $[i,j\rangle$ shift if $i$ and $j$ are both scalars.
Here we want to use supersymmetry to relate the large-$z$ behaviour of generic supersymmetric form factors
to that of form factors with legs $i$ and $j$ being both scalars.
This will then prove the  validity of the supersymmetric BCFW recursion relation for all supersymmetric form factors
in fashion similar to \cite{ahck}.

For supersymmetric non-chiral form factor $F(\lambda,\tilde\lambda,\eta_+, \tilde\eta^-)$, the $[i,j\rangle$ shift is
\bea && \hat{\tilde{\lambda}}_i(z) := \tilde{\lambda}_i + z \tilde{\lambda}_j ~, \qquad \qquad  \hat\lambda_j := \lambda_j - z \lambda_i ~, \nonumber\\
&& \hat \eta_{i,+} := \eta_{i,+} + z \eta_{j,+} ~, \quad~ \qquad \hat{\tilde{\eta}}_j^- = \tilde\eta_j^- - z \tilde\eta_i^- ~.
\eea
As in \cite{ahck}, we choose a particular
transformation where
\bea \bar Q_{\tilde\zeta} = \tilde\zeta_{\dot\alpha +} \bar Q^{\dot\alpha +} ~, \qquad Q_\xi = \xi_{\alpha}^{-} Q^{\alpha}_{-} ~,  \eea
where
\bea \tilde\zeta = {1\over [i~j]} \Big( - \tilde\lambda_i \eta_j + \tilde\lambda_j \eta_i \Big) ~,
\qquad \xi = {1\over\langle i~j\rangle} \Big(-\lambda_i \tilde\eta_j + \lambda_j \tilde\eta_i \Big) ~. \eea
One can check that their action on the fermionic coordinates $\eta_{k,+}, \tilde\eta_k^-$ is
\bea
e^{\bar Q_{\tilde\zeta}} \eta_{k,+} &:=& \eta_{k,+}' = \eta_{k,+} - \eta_{i,+} {[kj]\over[ij]} + \eta_{j,+} {[ki]\over[ij]} ~, \\
e^{Q_{\xi}} \tilde\eta_k^- &:=& \tilde{\eta}'^-_k =
\tilde\eta_k^- - \tilde\eta_i^- {\langle kj \rangle \over \langle ij \rangle} + \tilde\eta_j^- {\langle ki\rangle\over \langle ij\rangle} ~, \eea
and in particular $e^{\bar Q_{\tilde\zeta}} \eta_{i,+} = e^{\bar Q_{\tilde\zeta}} \eta_{j,+} = e^{Q_{\xi}} \tilde\eta_i^- = e^{Q_{\xi}} \tilde\eta_j^- =0$.
Since the form factor is invariant under  $\bar Q^+$ and $Q_-$ transformations, i.e.~$e^{\bar Q_{\tilde\zeta}} {\cal F} = e^{Q_{\xi}} {\cal F} = {\cal F}$ (see \eqref{holiday}), we conclude that
\bea &&
{\cal F}(\lambda_1,\tilde\lambda_1, \eta_{1,+}, \tilde\eta_1^-; \cdots;
\lambda_i, \hat{\tilde\lambda}_i, \hat{\eta}_{i,+}, \tilde\eta_i^- ; \cdots; \hat{\lambda}_j, \tilde\lambda_j, \eta_{j,+},  \hat{\tilde\eta}_j^- ; \cdots;\lambda_n,\tilde\lambda_n, \eta_{n,+}, \tilde\eta_n^-)
\nonumber\\ &=& {\cal F}(\lambda_1,\tilde\lambda_1, \eta_{1,+}', \tilde{\eta}'^-_1; \cdots;
\lambda_i,\hat{\tilde\lambda}_i, 0,0 ; \cdots; \hat{\lambda}_j,\tilde\lambda_j, 0,0 ; \cdots; \lambda_n,\tilde\lambda_n, \eta_{n,+}', \tilde{\eta}'^-_n) ~.
\eea
Thus,  we can always choose a supersymmetry transformation which sets  $i$ and $j$ to be scalars.
It is important to notice that under the $[i,j\rangle$ shift, the transformed $\eta_+'$ and $\tilde{\eta}'^-$ variables are  independent of $z$.
The large-$z$ behaviour of ${\cal F}(z)$ is therefore the same as that of the bosonic form factor with $i$ and $j$ being  scalars. This case was considered in the previous appendix, and shown to fall off as $1/z$ at large $z$. Hence the statement is also true for the shifted supersymmetric form factor ${\cal F}(z)$.
The proof illustrated above concerned the large-$z$ behaviour  of the full non-chiral super form factor, but a very similar one applies to  the form factor in chiral superspace,  since the latter is related to the former by a half-Fourier transform in superspace.

\section{Dual MHV rules \label{appendix-dualMHV}}

We recall that momenta and supermomenta for a massless particle are defined in terms of
dual momenta and supermomenta as \cite{dhks}
\bea x_i - x_{i+1} = \lambda_i \tilde \lambda_i ~, \qquad \Theta_i -
\Theta_{i+1} = \lambda_i \eta_i ~, \eea
where $x_{ij} := x_i - x_j$ and $\Theta_{ij} := \Theta_i -
\Theta_j$.

\begin{figure}[h]
\centerline{\includegraphics[height=6.3cm]{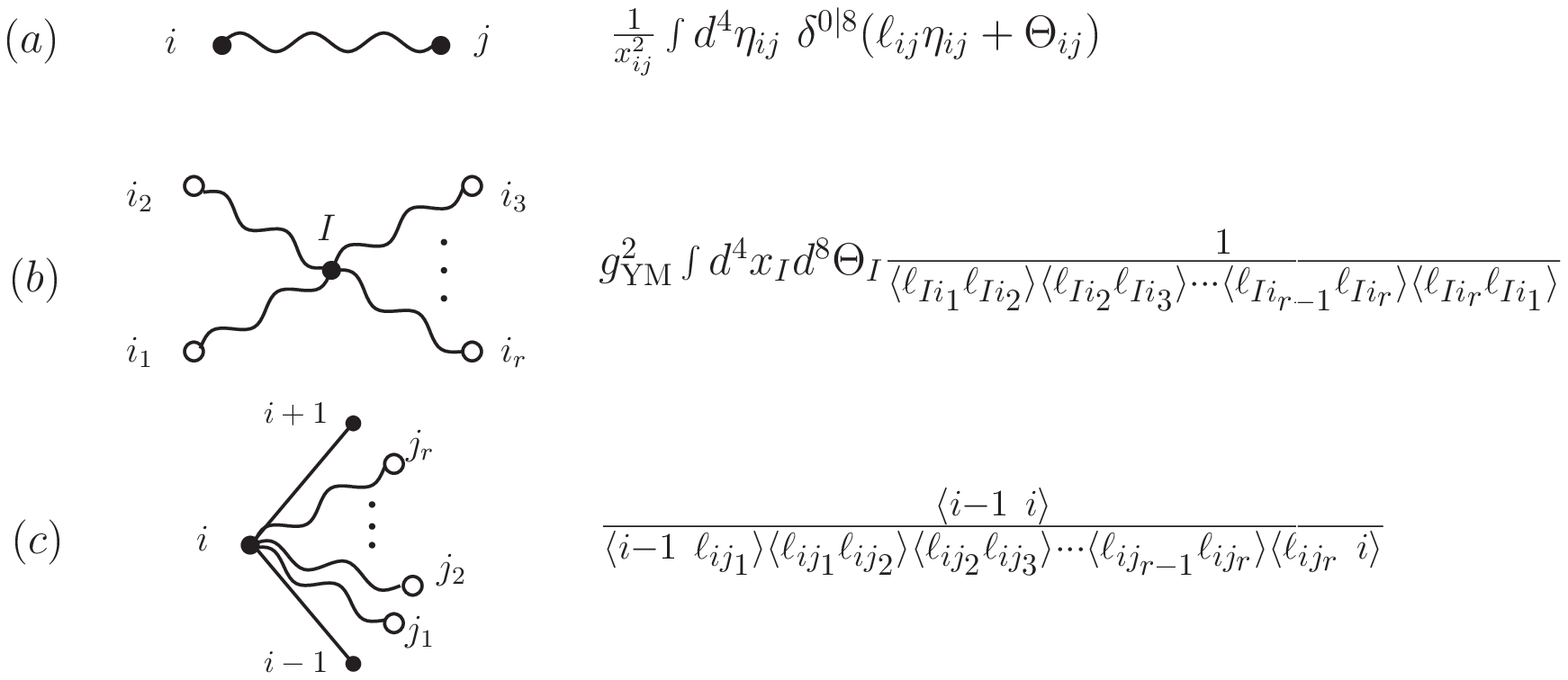} }
\caption{\it Feynman rules for dual MHV diagrams. (a) Propagator.
(b)  $r$-point internal vertex. (c) $r$-point external vertex.} \label{DualMHV-rules}
\end{figure}

The dual MHV diagram rules of \cite{bsty1} are summarised   in Figure
\ref{DualMHV-rules}.
%
%
%


In these rules, the off-shell continuation for spinors associated to internal legs, $|\ell_{ij}\rangle$, is defined as usual as \cite{csw}
\bea | \ell_{ij} \rangle := x_{ij}|\xi ] ~, \eea
where $|\xi]$ is an arbitrary reference spinor.

\begin{figure}[h]
\centerline{\includegraphics[height=2.5cm]{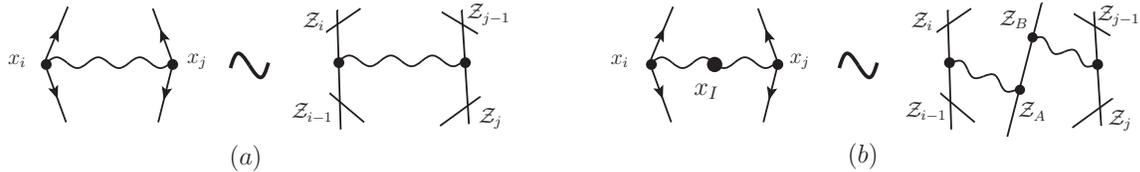} }
\caption{\it Dual MHV diagrams for (a) a NMHV tree amplitude, and  (b) a one-loop MHV amplitude,
and the  corresponding   momentum twistor diagrams.}
\label{DualMHV-STMT}
\end{figure}

For convenience, we recall here two simple applications of these
rules from \cite{bsty1}. A generic dual MHV diagram contributing to
an  NMHV tree amplitude is pictured in Figure
\ref{DualMHV-STMT}$(a)$. There are two boundary vertices and one
propagator. By applying the dual MHV rules of Figure
\ref{DualMHV-rules}, we get
\bea {\langle i\!-\!1 ~i \rangle \over \langle i\!-\!1 ~\ell_{ij}
\rangle \langle \ell_{ij} ~i \rangle} {\langle j\!-\!1 ~j \rangle
\over \langle j\!-\!1 ~\ell_{ij} \rangle \langle \ell_{ij} ~j
\rangle} {1\over x_{ij}^2} \int d^4 \eta_{ij} ~ \delta^{0|8}
(\ell_{ij} \eta_{ij} + \Theta_{ij}) ~, \eea
which can be easily translated in terms of the superconformal
invariant $R$-function  $R_{*;ij} := [*,i\!-\!1,i,j\!-\!1,j]$,
with
\be [ i,j,k,l,m ] \equiv { \delta^{(4)}(\langle i~j~k~l \rangle
\chi_m + {\textrm{cyclic terms}}) \over \langle i~j~k~l\rangle
\langle j~k~l~m \rangle \langle k~l~m~i \rangle \langle l~m~i~j
\rangle \langle m~i~j~k \rangle} ~. \label{R-inv} \ee
The reference momentum twistor is ${\cal Z}_* = (0, \xi,
0)$.

Similarly,  a generic dual MHV diagram for a one-loop MHV amplitude is depicted in Figure \ref{DualMHV-STMT}$(b)$ and is equal to
\bea &&  g_{\rm YM}^2 \int d^4 x_I d^8 \Theta_I {1\over \langle
\ell_{iI} \ell_{Ij} \rangle \langle \ell_{Ij} \ell_{iI} \rangle } \
{\langle i\!-\!1 ~i \rangle \over \langle i\!-\!1 ~\ell_{iI} \rangle
\langle \ell_{iI} ~i \rangle} {\langle j\!-\!1 ~j \rangle \over
\langle j\!-\!1 ~\ell_{Ij} \rangle \langle \ell_{Ij} ~j
\rangle}  \nonumber\\
&& {1\over x_{iI}^2} \int d^4 \eta_{iI} ~ \delta^{0|8} (\ell_{iI}
\eta_{iI} + \Theta_{iI}) {1\over x_{Ij}^2} \int d^4 \eta_{Ij} ~
\delta^{0|8} (\ell_{Ij} \eta_{Ij} + \Theta_{Ij}) ~. \eea
In terms of momentum twistor variables this becomes
\cite{Bullimore:2010pj},
\bea g_{\rm YM}^2 \int d^{3|4} {\cal Z}_A \wedge d^{3|4} {\cal Z}_B
~ [* , i\!-\!1 , i , A , B'] [* , j\!-\!1 , j , A , B''] ~, \eea
where $(x_I,\Theta_I) \sim {\cal Z}_A \wedge {\cal Z}_B$ and
\bea B' = (A,B) \cap (*,j\!-\!1,j) ~, \qquad B'' = (A,B) \cap
(*,i\!-\!1,i) ~. \eea
%


\end{document}